\documentclass[journal]{IEEEtran}

\usepackage{amsthm,amsmath,amstext,amsfonts,amssymb,mathrsfs} 
\usepackage{graphicx,subfig,algorithm,algorithmic,color}

\theoremstyle{remark}
\newtheorem{remark}{Remark}

\usepackage{multicol}

\usepackage{hyperref}

\newcommand{\g}{\gamma}

\title{Estimating  Phase Duration for SPaT Messages}

\author{Shahana Ibrahim, Dileep Kalathil, Rene O. Sanchez and Pravin Varaiya,~\IEEEmembership{Life Fellow,~IEEE}
\thanks{Shahana Ibrahim and Dileep Kalathil are with the Department of Electrical and Computer Engineering, Texas A\&M University. Rene O. Sanchez is with Sensys Networks, Berkeley, California.   Pravin Varaiya is with  the Department of Electrical Engineering and Computer Sciences, University of California, Berkeley. Corresponding author email: { dileep.kalathil@tamu.edu}}
\thanks{This research was supported by the California Department of Transportation UCTC Award 65A0529 TO041 and Sensys Networks, Inc.}
}

\begin{document}
\maketitle
\markboth{Submission for IEEE Transactions on Intelligent Transportation Systems} 
{Ibrahim \emph{et al.}}

\begin{abstract}
A SPaT (Signal Phase and Timing) message describes for each lane the current  phase at a signalized intersection together with an estimate of the residual time of that  phase.  Accurate SPaT messages can be used  to construct a speed profile for a  vehicle that reduces its fuel consumption  as it approaches or leaves an intersection.  This paper presents  SPaT  estimation algorithms at an intersection with a semi-actuated signal, using real-time  signal phase measurements. The algorithms are evaluated using high-resolution data from an intersection in Montgomery County, MD.  The  algorithms can be readily implemented at signal controllers.  The study supports three findings.  First, real-time information dramatically improves the accuracy of the prediction of the residual time compared with prediction based on historical data alone.  Second, as time increases the prediction of the residual time may increase or decrease.  Third,
as  drivers differently weight errors in predicting `end of green' and `end of red', drivers on two different approaches may prefer different estimates of the residual time of the \textit{same} phase.
\end{abstract}
	

\section{Introduction}
A SPaT (Signal Phase and Timing) message describes the current  phase at a signalized intersection, together with the residual time of the   phase, for every lane (hence every approach and movement) of the intersection.  The estimate is periodically broadcast by the intersection, say once per 100ms. For a fixed-time controller the SPaT information is definitive; the challenge is for an actuated controller for which only an estimate of the residual time can be given, and for which  the SPaT message data elements include StartTime of the phase, its MinEndTime, MaxEndTime, LikelyTime, Confidence (in the LikelyTime) and NextTime (when this phase will next occur).

A SPaT message is used together with a MAP message, which describes the physical geometry of one or more intersections.  A vehicle approaching or leaving the  intersection, with knowledge of its own position and speed and MAP information, can take the residual time of the current phase from the SPaT message to calculate a speed profile that reduces stop-and-go driving and idling \cite{GAO-15-775}. No intersection in the U.S. today broadcasts SPaT or MAP messages, and very few cities have a Traffic Management Center that receives phase information from all its intersections.  But in our experience, it is  inexpensive to collect and process phase information locally at each intersection, and  the significant interest and activity in the automotive and ITS communities devoted to standardization of SPaT and MAP message formats \cite{J2735, PATH_Spat} suggest that intersections that deliver SPaT may become common.

Several  studies  offer  `eco-friendly' speed advice based on SPaT and simulate its benefits.    A model predictive control (MPC) is used in \cite{vahidi_spat_2011} to construct a vehicle trajectory that traverses a sequence of intersections without stopping at a red light, knowing the  SPaT sequence in advance.  The  speed advice in \cite{rakha_spat_2011}  is based on SPaT messages from the upcoming intersection and  the expected reduction in emissions and fuel consumption is evaluated using a vehicle dynamics model.  A  velocity planning algorithm for eco-driving through a signalized corridor is simulated in \cite{barth_spat_2011}.  These  studies report fuel savings in arterial driving ranging between 12 and 47 percent.  This wide variability in fuel savings estimates may be due to differences in the underlying vehicle simulation models.    
	
On the other hand,  the Glidepath field experiment found  drivers who got speed advice saved 7 percent, whereas SPaT-based automatic speed control saved 22 percent of fuel, relative to an uninformed driver \cite{Glidepath}. The Glidepath experiments suggest that accuracy in the SPaT estimate is essential to maximize the fuel savings that automation can achieve.  Another field experiment \cite{xia2012field} reports comparable fuel savings. 

SPaT messages may also improve safety by preventing rear-end accidents due to errors in driver predictions of the phase duration and also because drivers can know which conflicting approaches have the right of way.
For a review of  use cases for  SPaT  see \cite{PATH_Spat};  \cite{AG_spat} gives a European perspective.
	
\subsection{Related Work}
	
	A few studies propose  SPaT estimation  based on  noisy  measurements of signal  phase.    The  approach in 
	\cite{fayazi_spat_2015} estimates the cycle length, phase  durations, and the cycle start time for several intersections along a segment of Van Ness Avenue in San Francisco.  Only  fixed-time signals are considered.  
	The data consists of  samples of GPS position and speed taken  every 90s or 200m from  4,300 bus runs over this segment for one month.   However, to estimate the red duration at an intersection, only between 40 and 350 samples that occurred right before and right after an intersection were found to  be usable.   These few  samples were ``aggregated'' to estimate the duration of red. The accuracy of the estimates is unimpressive, with absolute errors up to 6s for a red duration of 36s \cite[Table I]{fayazi_spat_2015}. (By way of  comparison, we report in Figure \ref{d4mae} below the mean absolute error for an actuated signal   between 0.5 and 3s for an average phase duration of 38s.) Since the signal timing parameters are fixed and available from the San Francisco Transportation Authority, the one month-long data collection and processing effort to estimate these parameters  seems misspent.  A follow-on study that simultaneously estimates the waiting time spent by the bus in queue shows a significant
	improvement in the SPaT estimate \cite[Figure 16]{fayazi_spat_2016}.
	The approach to estimating  parameters for fixed-time signals by associating speed measurements of ``floating cars'' to the signal status was earlier  exploited in \cite{ban2009delay,wang2012traffic,protschky2015learning}.  
	
	The scheme described in \cite{signalguru}  engages  several cooperating drivers with smartphones    to locate (detect) the signal light  at an intersection and predict its phase duration.  Much effort is spent to  detect  the signal head and identify the signal color. (In the two reported scenarios, the signal head misdetection rates were  7.8 and 12.4 percent, meaning that in one out of ten intersections on average, the signal head was not detected or some other object was mistaken for a signal head.)  The two remaining tasks are (1) estimating the phase durations and (2) `synchronization' or locating the  current time within the current cycle or phase, so the  vehicle can figure out the residual time of the phase. In case of a fixed-time signal with known timing plan, the phase durations are known, and synchronization is equivalent to determining the time of a phase transition from (say) green to red.   For an actuated signal the phase duration varies from one cycle to the next, and three machine learning algorithms are tested to predict the duration of the next phase from  signal phase history; however, the best prediction based on  the five previous phases and  cycle lengths is only slightly better than taking the next phase duration to be the same as the last duration.  The authors do not discuss how phase duration estimates made by  vehicles at earlier times are transferred to the  vehicle that is making the current prediction.

	The two studies \cite{fayazi_spat_2015, signalguru} spent much effort in collecting and processing noisy measurements of signal phase.  By contrast, city transportation agencies and  auto companies  obtain the signal phase data directly from the signal controller. Cities may invest in SPaT devices to improve mobility generally and to provide priority to public transportation \cite{nocoe}. Auto companies
devote their effort to improving SPaT algorithms and  designing interfaces to present SPaT messages to drivers.

The study \cite{BMW_Spat} relies on  data collected from  traffic intersections to make a probabilistic SPaT prediction. Phase duration data  is used to compute its empirical frequency distribution at the beginning of each cycle. Then for each second $i$ in the cycle, the prediction $\gamma_{k}(i)$  of phase $k$ takes the value  $\g_k(i) =G $ or $R$ if with (say) 80 percent `confidence'  phase $k$ is green or red  at second $i$, whereas $\g_k(i) = M$ means the phase may be green or red. But to figure out the residual time of the current phase, a vehicle also needs to know the  time within the cycle, although \cite{BMW_Spat} does not mention the need to broadcast this time.  Even with this time, however, the prediction  in  \cite{BMW_Spat}  may  be uninformative. Of course the `80\% confidence' predictions may be incorrect; more importantly, they leave a lot of uncertainty that will grow if we require a higher   confidence level. The deeper problem with compressing the raw  data  into  frequency distributions  as in \cite{BMW_Spat}  is that the frequency distribution obscures the residual times of the phase, which is what the SPaT message is supposed to provide and  what vehicles  need to design their speed profile.

 The brief description in \cite{Audi_Spat} does not explain how its SPaT estimate is derived but states that it is displayed to the driver via a ``countdown clock on the dashboard.''
	The website (conectedsignals.com) declares  that it combines real-time signal data (presumably obtained from the Traffic Management Center), with GPS location, and speed limit information, to predict upcoming traffic signal behavior and deliver it  via cellphones.  
	
Several papers study predicting  vehicle flow at arterial roads and using the prediction for adaptive signal control  \cite{sun2012vehicle} \cite{chen2016improved} \cite{coogan2017traffic}. Our focus on predicting the phase duration at traffic intersections is quite different. 
	
\subsection{Contributions of this paper} We develop   algorithms to estimate the residual duration of every phase for two semi-actuated intersection in Montgomery County, MD. Detailed description of the algorithms for one intersection is provided; the study of the second intersection  suggests that the same procedure can be applied elsewhere.
Direct measurements from the intersection provide the ground truth   used to evaluate the   algorithms.  The algorithms predict the  times for all future phase transitions, based on previous phase measurements and on the real time information that locates the current time within the current phase.  For actuated signals, conditioning the prediction on this real time information greatly reduces the prediction error.  To our knowledge this is the first paper to use this information, which is  available at the signal controller. Perhaps  surprising is the finding that for semi-actuated signals, as time increases, the estimate of the residual phase duration may increase or decrease, posing a challenge to construct fuel-minimizing speed profiles.  
	Another contribution stems from the  observation that for a driver the best SPaT estimate is the one that minimizes the driver's own loss function.  For example if, as seems likely, drivers differently weight errors in predicting `end of green' and `end of red', drivers on two different approaches would prefer different estimates of the same phase transition, since `end of green' for  one approach to the intersection would be the `end of red' for  the other.  This suggests that multiple SPaT estimates should be created and broadcast.
	
The paper is organized as follows.  Section \ref{sec_measure} describes the intersection site and the measurement system.  Section \ref{sec_problem} formulates the SPaT estimation problem. Sections \ref{sec-residual} describes the prediction algorithms and the evaluation of their performance. Section \ref{sec-loss} describes prediction as the minimization of a loss function.  Prediction of the duration of  future phases is described in Section  \ref{sec:otherphases}. Section \ref{sec:site2} describes the evaluation of the algorithms using the data from another intersection.   Section \ref{sec-conc} collects the main conclusions.

\section{Measurement site} 
\label{sec_measure}
\begin{figure}[h!]
\centering
\includegraphics[width=0.45\textwidth]{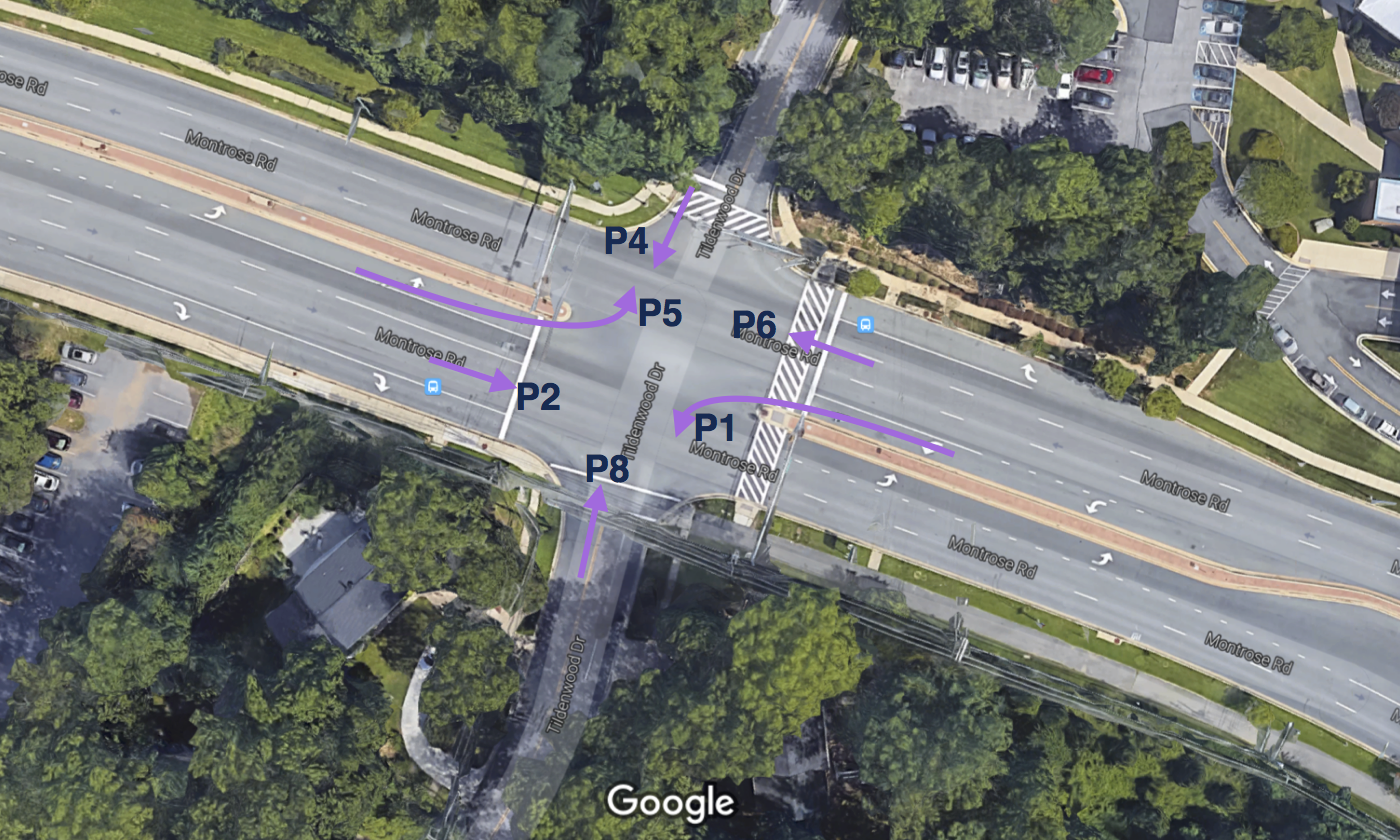} 
\caption{The intersection at Tildenwood Dr. and Montrose Rd.} 
\label{fig1}
\end{figure}
 
\begin{figure}[h!]
		\centering
		\includegraphics[width=0.45\textwidth]{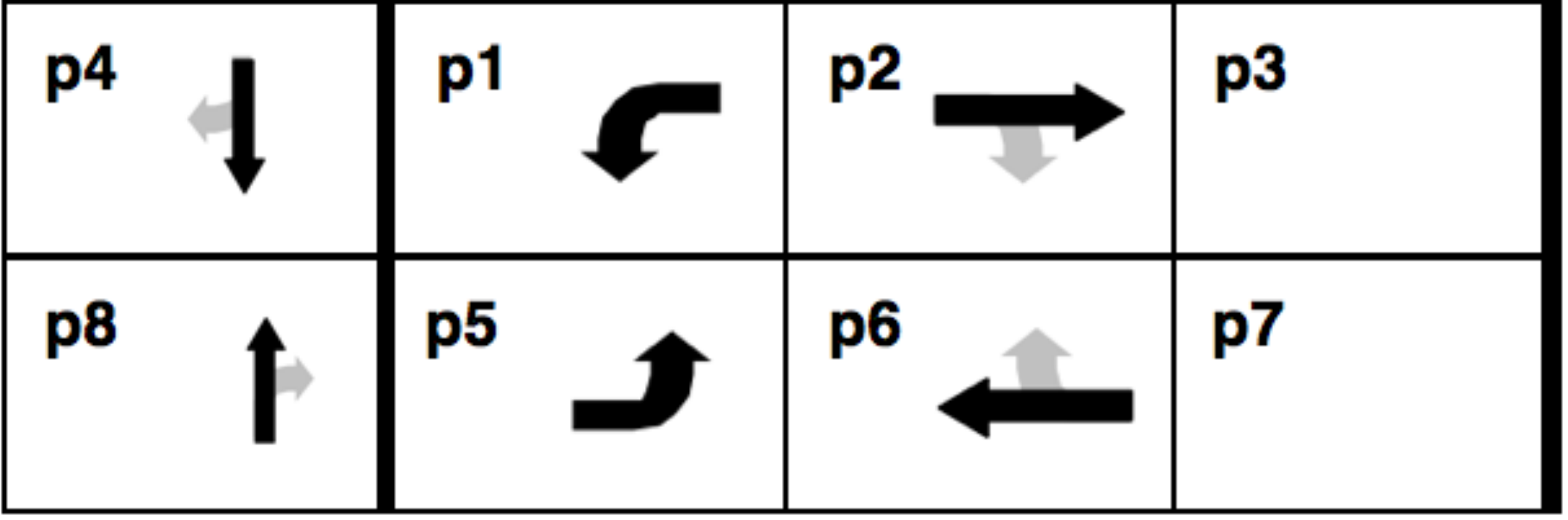} \quad
		\caption{The dual ring  diagram; phases $p_3$ and $p_7$ have zero duration. The thicker vertical lines are the two barriers. } 
		\label{fig2}
	\end{figure}

Figure \ref{fig1} shows the intersection at Tildenwood Drive and Montrose Road in Montgomery County,  MD, the first measurment site we consider (the second measurement site is described in Section \ref{sec:site2}).  The  same figure also indicates the six  phase movements permitted at this intersection.  The movements are  arranged in the dual ring and barrier structure  of Fig. \ref{fig2} (The ring structure is described in \cite{stm}.)  Ring 1 comprises phases $p_4$, $p_1$, $p_2$, and ring 2 comprises phases $p_8, p_5, p_6$; phases $p_3 ,p_7$ have zero duration.  The thicker vertical lines are the two barriers.  Across Montrose Rd., to the east of the intersection, is the pedestrian cross-walk  which is actuated in every cycle at the beginning of phases $p_4$ and $p_8$.   

	The intersection is equipped with magnetic vehicle detectors at the stop bar, at advance locations, and  in the departure lanes.  The latter permit an accurate count of turn movements ($p_1$ and $p_5$).  In addition, the current signal phase is obtained from the controller every 100ms.  All measurements are time stamped with a common clock with a 10ms accuracy.  These detectors are for measurement only; the controller itself relies on different detectors for signal control.
	
	In summary, the  data consists of the time series of vehicle detections and signal phase at a time resolution of 10ms.  The data is sent wirelessly to an access point (AP) located at the controller, from where it is sent via cellular connections to the traffic management center and to our server.  The data are obtained courtesy of Sensys Networks, Inc.   The analysis here uses two
		months of data from September and October 2016.  \textit{Only phase data is used in this study}; a future paper will report on the additional predictive power provided by vehicle detection data.

	The intersection is regulated by a semi-actuated, coordinated controller.  The cycle length is fixed by the timing plan at $L =$ 100, 110 or 120s, depending on time of day and day of week.  The cycle  is divided into nominal durations for each phase; the controller modifies  these durations in each cycle depending on vehicle detections. Phase $p_2$ or $p_6$ is the coordination phase, implying thereby that this phase is the last one in the cycle to receive its allocation of green time.  (The operation of actuated controllers is described in \cite{stm}.)  The main direction of traffic is East-West (Tildenwood) and   this direction has no detectors for signal control.  There are detectors in the secondary North-South (Montrose) direction.  If few vehicles are detected in the secondary direction, its green duration (phases $p_4$, $p_8$) is shortened and the time saved thereby is added to the  duration of the  phases in the main direction, $p_2$ and $p_6$.  Vehicles making left turn movements (phases $p_1$, $p_5$) are also detected, and their green duration is also reduced if fewer vehicles are detected in the turn pockets, see Figure \ref{fig1}.
	
	Thus if we  denote by $d_i$ the duration of phase $p_i$, we see that all of these durations may vary from one cycle to the next, while maintaining some identities:
	\begin{eqnarray}
	d_4 + d_1 + d_2  &=& d_8 + d_5 + d_6 = L ,\label{eq1} \\
	d_1 + d_2 &=& d_5 + d_6, \label{eq2} \\
	d_4 &=& d_8 .\label{eq3}
	\end{eqnarray}
	Equation \eqref{eq1} recognizes $L$ as the cycle length;  \eqref{eq2} and \eqref{eq3} are implied by the two barriers shown in Figure \ref{fig2}.

\section{The SP{\normalfont a}T Problem}  
\label{sec_problem}
	
\begin{figure} [h!]
\centering
\includegraphics[width=0.50\textwidth]{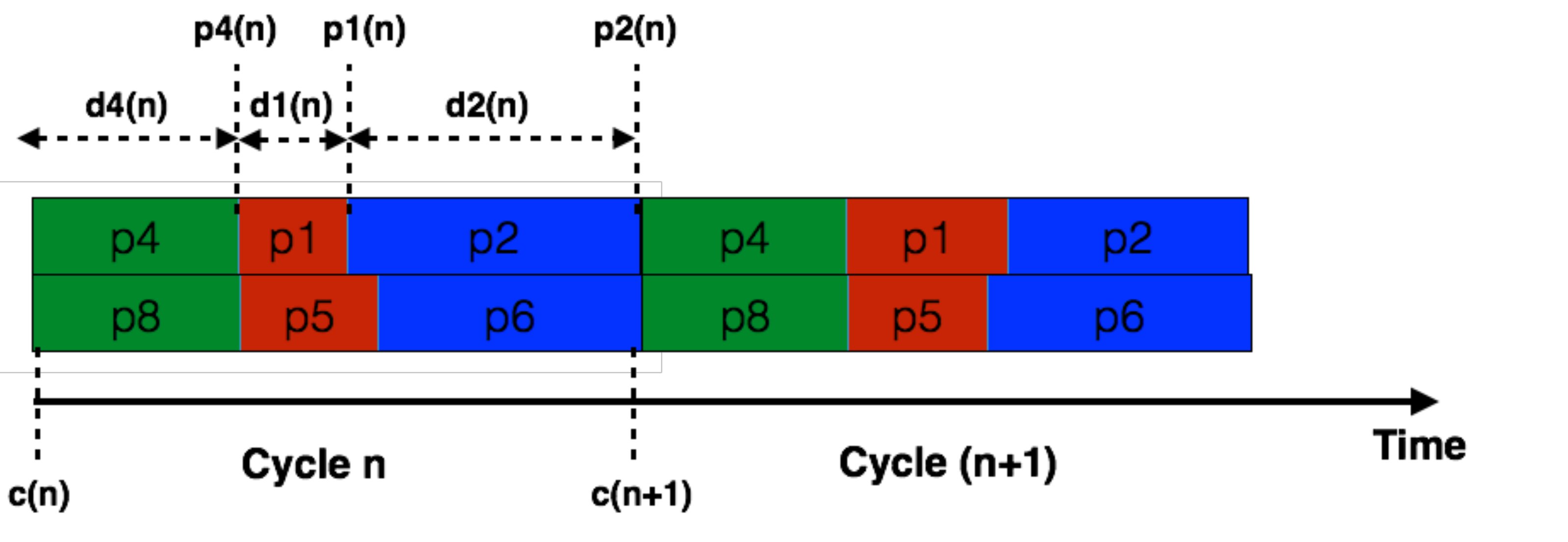}
\caption{Variables used to define the SPaT estimation problem.} \label{fig3} 
\end{figure}

We use Figure \ref{fig3} to define the SPaT estimation problem.  Time is in seconds.  The figure shows two cycles  $n$ and $n+1$, each of length $L$, starting at times $c(n)$ and $c(n+1)$;  $d_4(n), d_1(n) \cdots$ is the duration and $p_4(n), p_1(n), \cdots$ is the end time of phase $p_4, p_1, \cdots$ in cycle $n$;   so if $t$ is the current time in cycle $n$ during phase $p_4, p_1, \cdots $, then $p_4(n) - t, p_1(n) - t, \cdots $ is the residual time  of the phase that is included in the SPaT message.  Observe that from the phase end times one can calculate the phase durations, e.g. $d_1(n) = p_1(n) - p_4(n)$, etc.  Conversely, from the phase durations one can calculate their end times.
	
\textbf{The SPaT problem:} Let $I(t)$ be the information  about previous phases available at   time $t \in  [0,L]$ during cycle $n$.  The  problem is to predict the residual times $p_k(m) -t$ of all phases $k$ for all future cycles $m = n, n+1, \cdots$, given $I(t)$.   
	
\begin{figure} [h!] 
		\centering
		\includegraphics[width=0.50\textwidth]{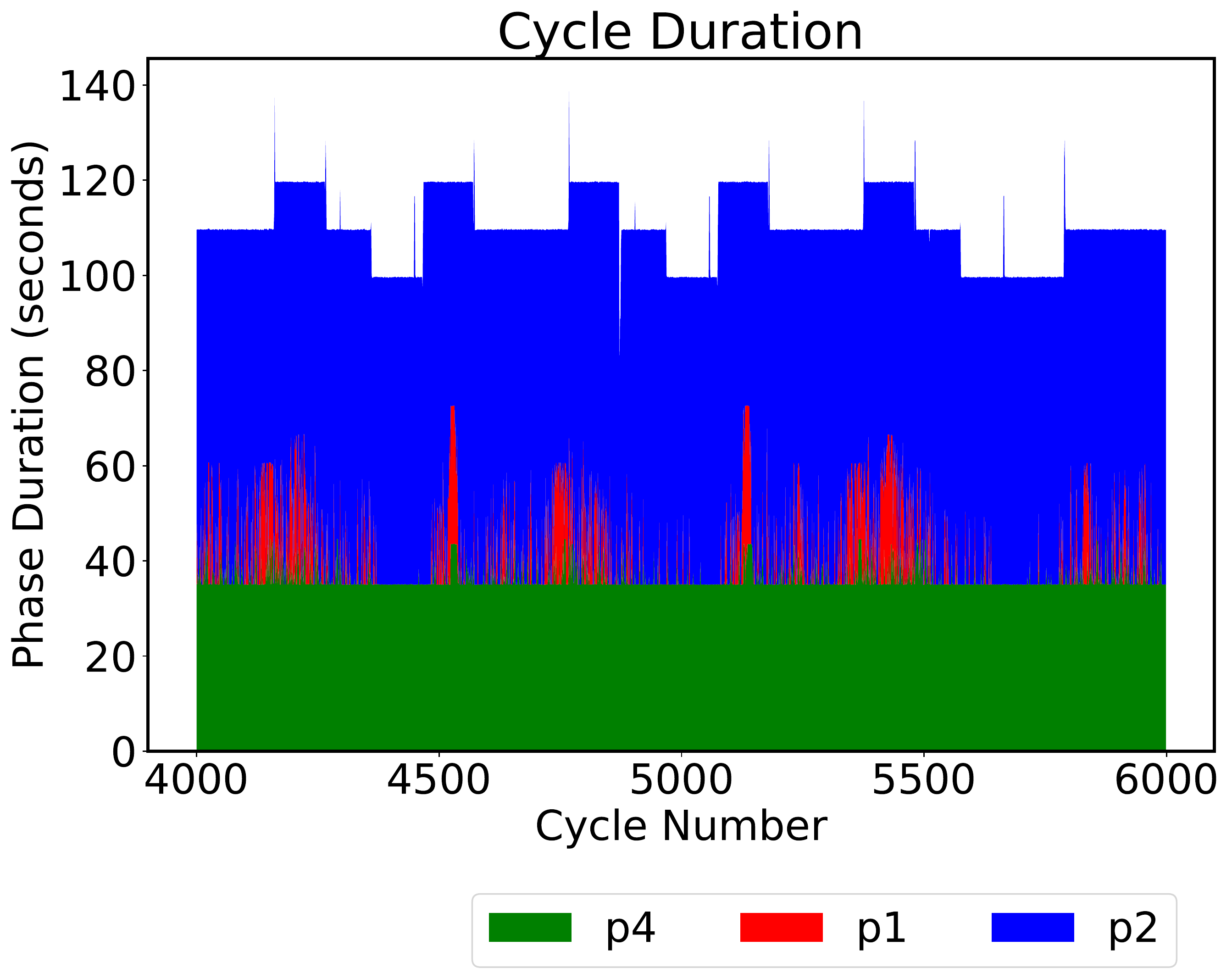}
		\caption{Variation in Phase durations over 2,000 cycles.} \label{fig4} 
\end{figure}

Our study uses data for 36,000 cycles from September  and October 2016. The phase data for a sample of 2,000 cycles (about 3 days) is shown in  Figure \ref{fig4}.  The plot  is similar to Figure \ref{fig3}: the difference is that the plot is rotated 90 degrees, the 2,000 cycles are `stacked' horizontally, and only phases in ring 1 are shown (there is a similar plot for phases in ring 2).   The $x$ coordinate is the cycle number; the $y$ coordinate is the time in seconds during that cycle.  The length of a cycle is 120, 110 or 100s, as determined by the timing plan.   Every cycle starts in  phase $p_4$ (and $p_8$); its duration is $d_4$ and it is colored green as in Figure \ref{fig3}. The minimum value of $d_4$ is the pedestrian  clearance time of 36s; the duration $d_4$ (and $d_8$) is extended by 5s each time an additional vehicle is detected (Variation of $d_{4}$ is difficult to see in the above figure, but will be clear in later plots). Phase $p_4$ is followed by the left turn phase $p_1$ (and $p_5$) colored red, and lasting $d_1$.  Duration $d_1$ depends on the detection of left turn vehicles, and each new detection triggers an extension of 5s.  $d_1$ is  zero in many cycles, when no left turn vehicles are detected.  The  cycle ends in the coordination phase $p_2$ (and $p_6$), colored blue, and lasts for time $d_2$.  (The infrequent `spikes' in $d_2$ occur when a change in the cycle length dictated by the timing plan is accommodated  over several cycles.)  Large values of $d_4$ and $d_1$ occur only during the AM and PM peaks.  The 4s yellow and 1.5 or 2s red signal phases are included in the green duration.

We now  present several algorithms for phase duration prediction for SPaT.

\section{Phase Duration Prediction}
\label{sec-residual}
\begin{figure*}[!t]
\centering
\subfloat[]{\includegraphics[width=0.32\textwidth]{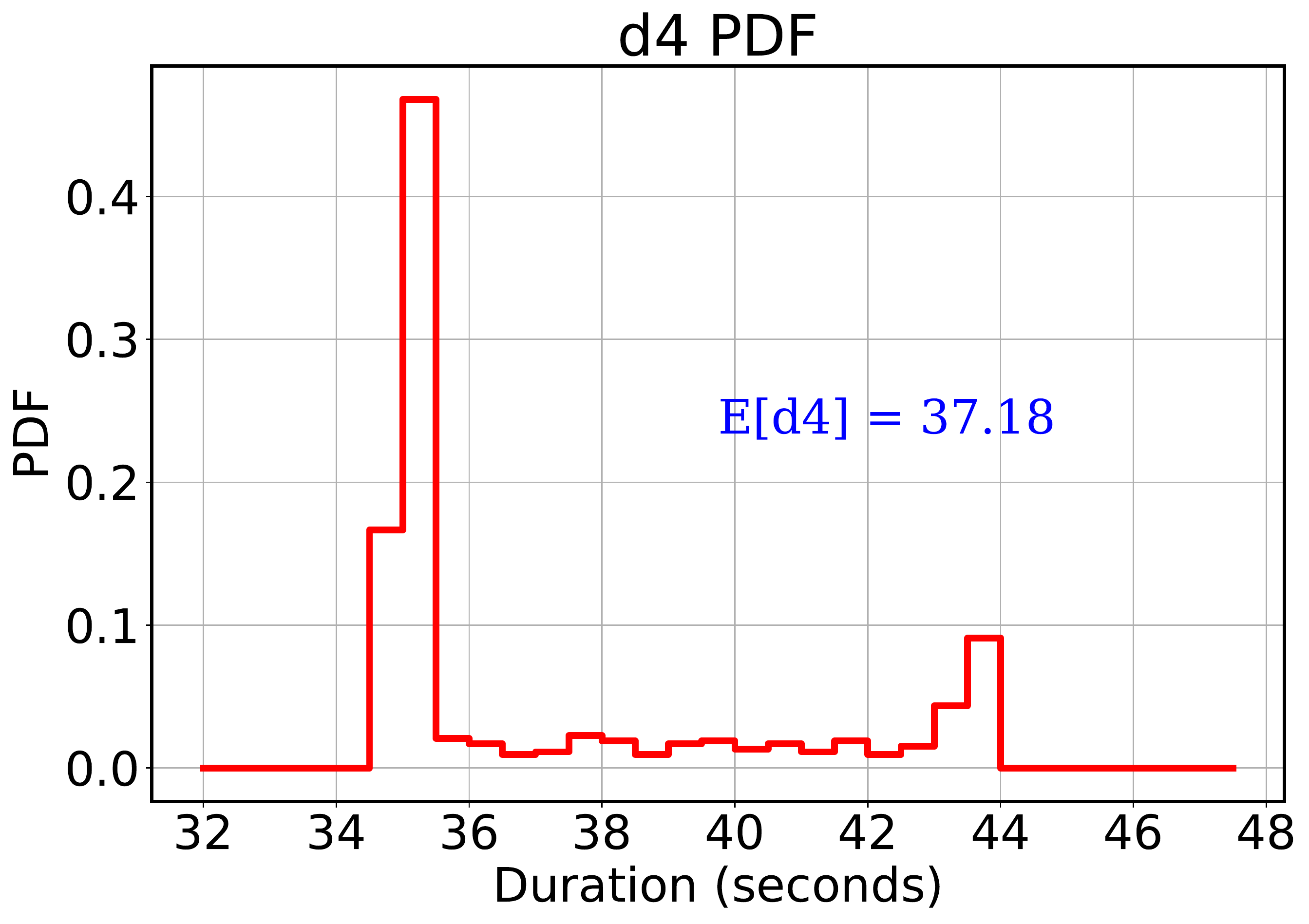}
\label{d4pdf1}}
\hfil
\subfloat[]{\includegraphics[width=0.32\textwidth]{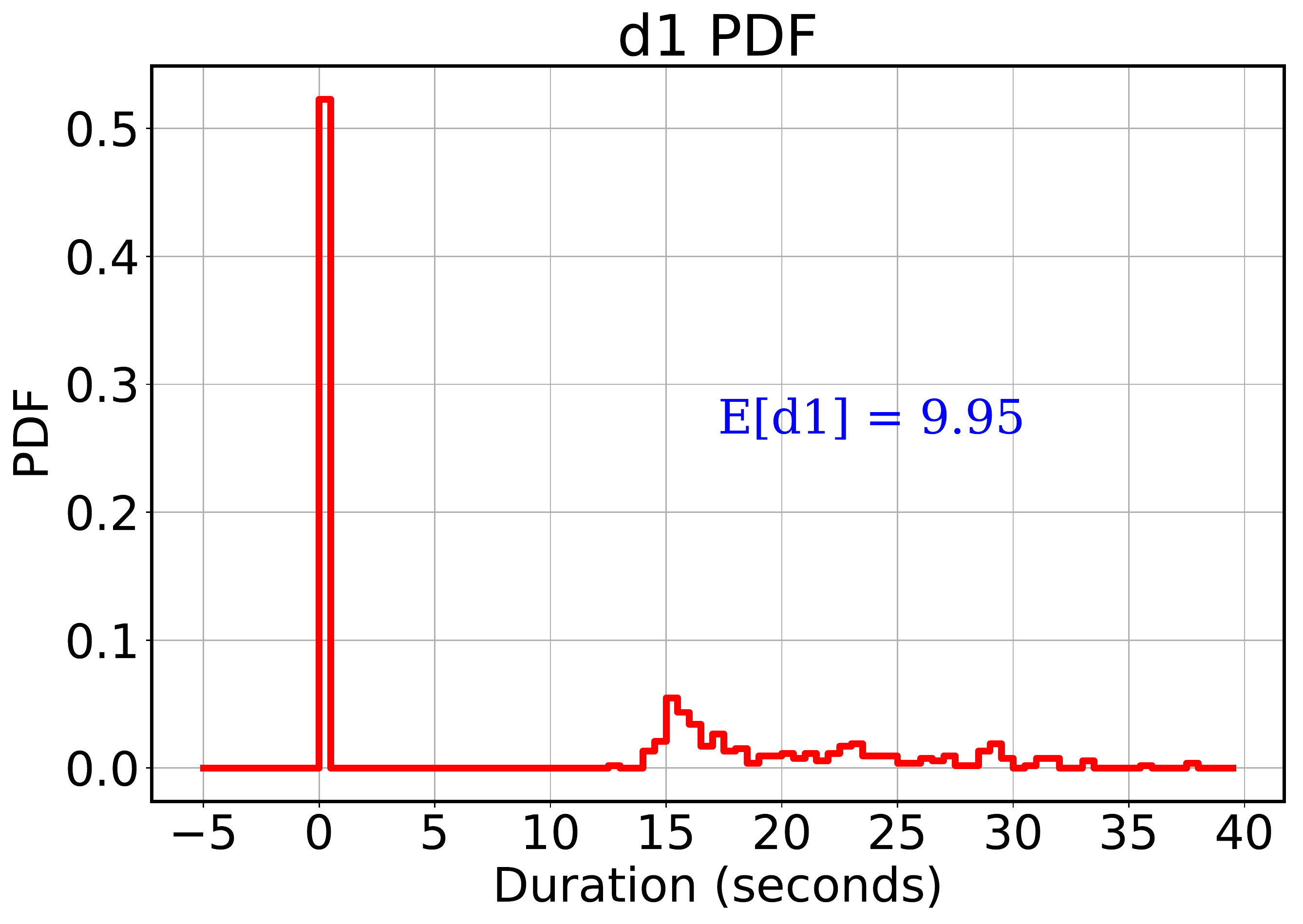}
\label{d1pdf1}}
\hfil
\subfloat[]{\includegraphics[width=0.32\textwidth]{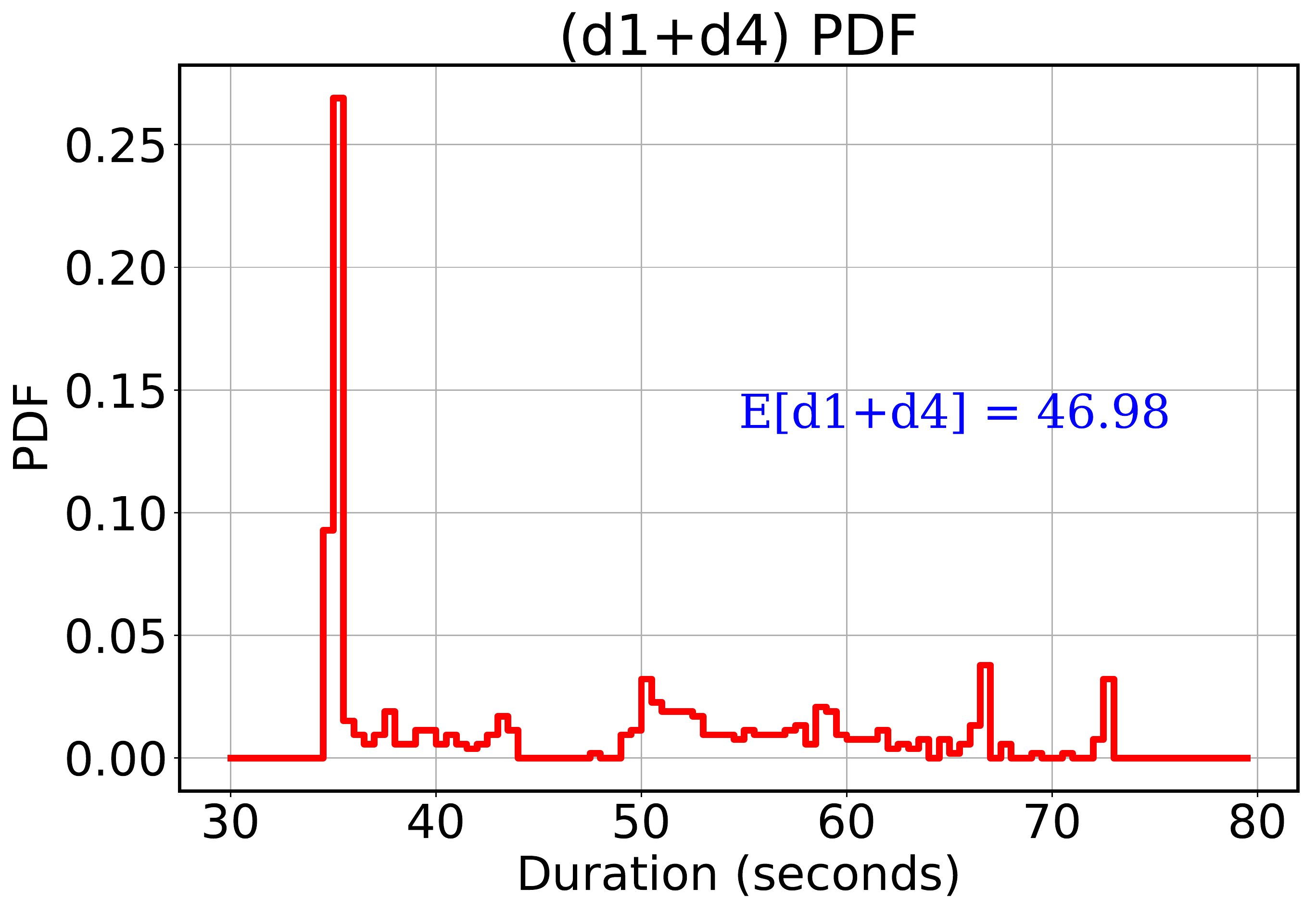}
\label{d4d1pdf1}}
\caption{PDF of $d_{4}, d_{1}$ and  $(d_{4} + d_{1})$.} 
\label{pdfplots}
\end{figure*}

As in \eqref{eq1}-\eqref{eq3} and Figure \ref{fig3}, let  $d_4, d_1, d_2$ denote the  duration of phase $p_4, p_1, p_2$.  (The treatment of phases $p_8, p_5, p_6$ is  analogous.) Since in each cycle, each phase is actuated for a contiguous interval of time, it is easy to calculate the histograms or empirical probability distributions (pdf) of the durations from the raw data of Figure \ref{fig4}. In the following, we only consider the phase duration data for cycles whose length is 120 seconds, as it makes no sense to aggregate durations  from cycles with different length.  Since $L = 120$ and $d_2 = 120 -(d_4 + d_1)$ (by \eqref{eq1}),  it is enough to calculate the pdfs of $d_4, d_1, d_4+d_1$. Since $d_4$ and $d_1$ may be dependent, the pdf of $d_4 + d_1$ cannot be calculated from the pdfs of $d_4$ and $d_1$.  The calculated pdfs are plotted in Figure \ref{pdfplots}.  Separate pdfs must be calculated for cycles of length 100 and 110 seconds.

In the following we propose two approaches for predicting the phase duration using this empirical pdf: a conditional expectations based prediction and a confidence based prediction. The only real-time information  used  is the phase and the amount of elapsed time of the phase at   time $t$ in the cycle when the prediction is made.
	
\subsection{Conditional expectation  based prediction} 
\label{sec-conditional}
If the prediction is made at the start of the cycle, $t = 0$,  nothing is known except the unconditional pdfs of  the phase durations in Figure \ref{pdfplots}, and so   a reasonable prediction for the durations is their expected values.  These values, $\mathbb{E}[d_4], \cdots$,  are inserted in the plots of Figure \ref{pdfplots}.  Now consider the prediction of the residual duration of $d_4$ at some later time $t$ in the cycle.  We see from the pdf of $d_4$ that if $t> 36$ and if  $d_4$ is still actuated at $t$, i.e.  $d_4 > t$, a  better prediction at $t$ would be  the expected value of $d_4$, conditioned on the event $\{d_4 > t\}$.  We define this prediction $\hat{d}_{4}(t)$ as
\begin{equation}
\hat{d}_{4}(t) =  \mathbb{E}[d_4~|~ d_4 > t].
\end{equation}

\begin{figure*}[t!]
		\centering
		\subfloat[]{\includegraphics[width=0.31\textwidth]{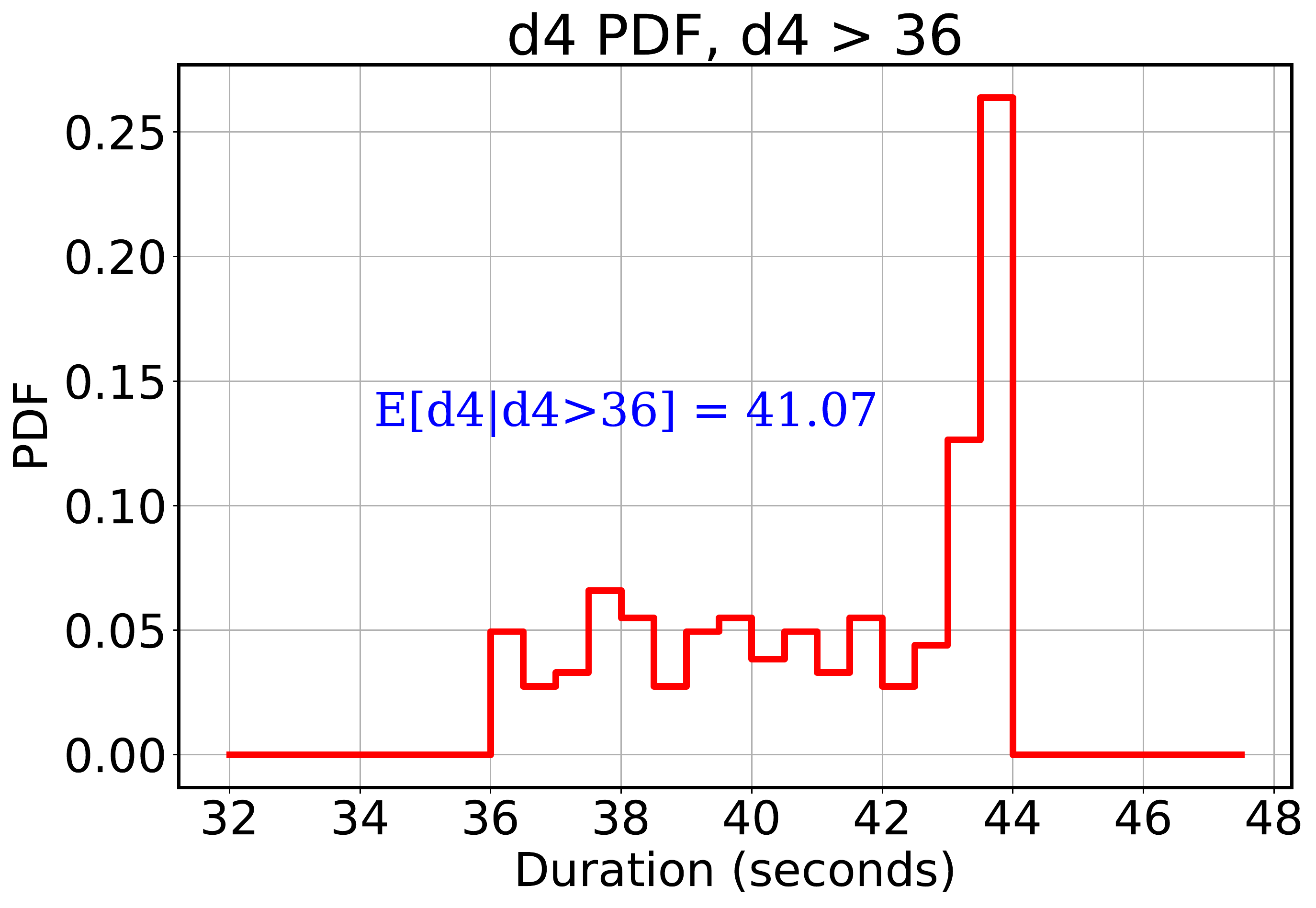} 
			\label{d4cep1}}
		\hfil
		\subfloat[]{\includegraphics[width=0.31\textwidth]{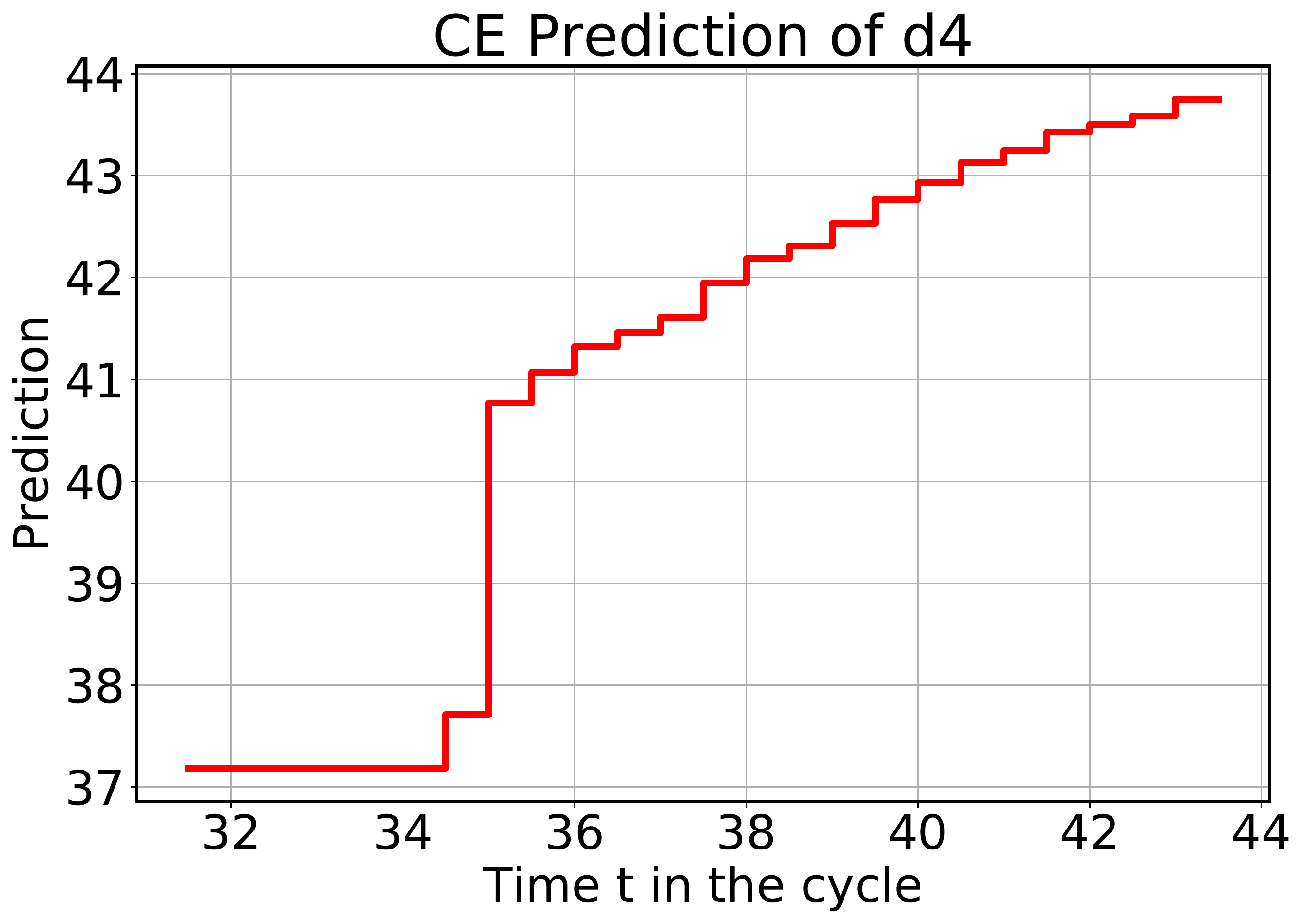}
			\label{d4cep2}}
		\hfil
		\subfloat[]{\includegraphics[width=0.31\textwidth]{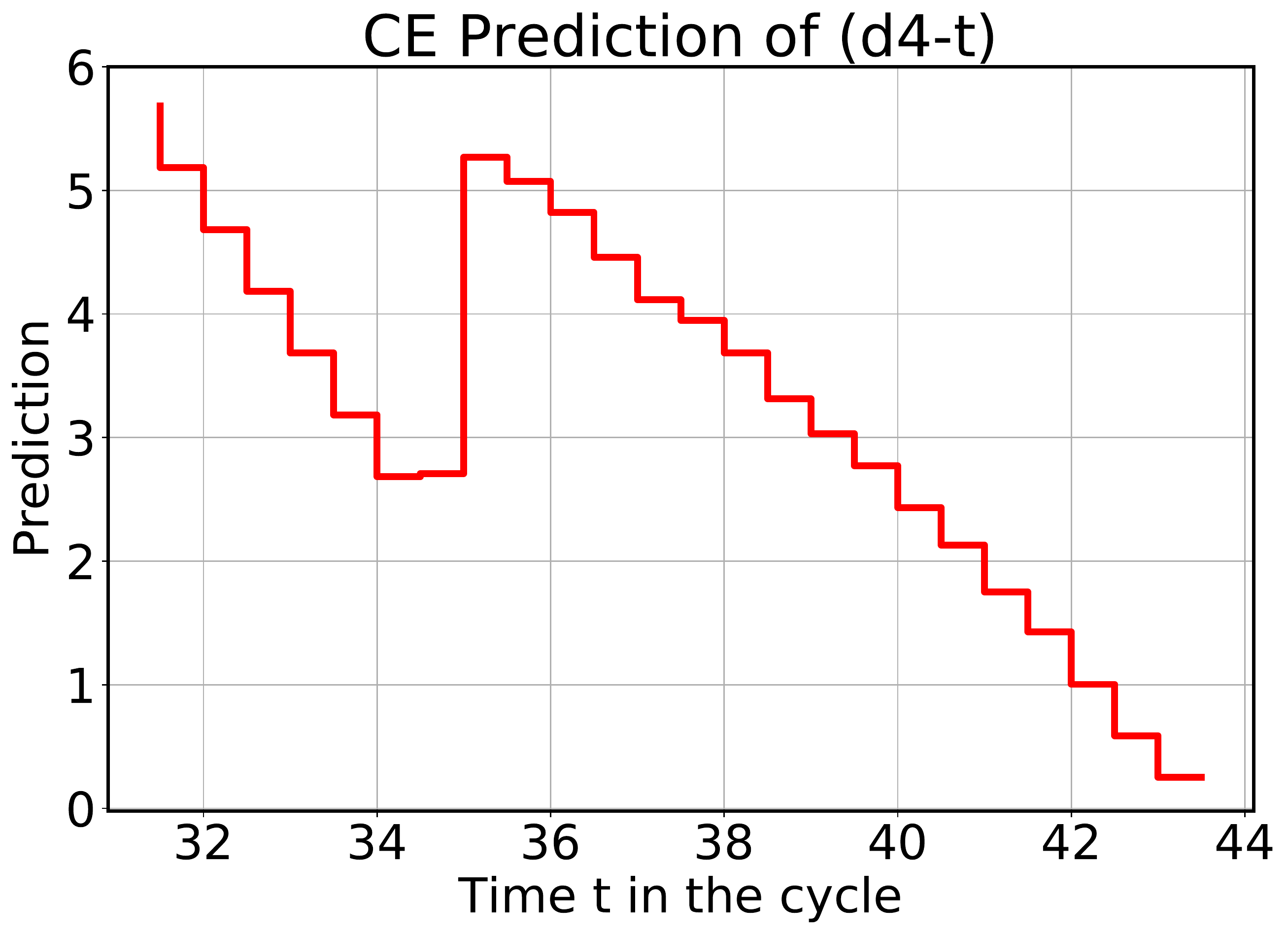}
			\label{d4cep3}}
		\caption{Fig. \ref{d4cep1} shows the conditional pdf of $d_4$, conditioned on $\{d_4 > 36\}$.  Fig. \ref{d4cep2} shows the conditional expectation based prediction of $d_4$. Fig. \ref{d4cep3} shows the conditional expectation based prediction of the residual time $(d_4 - t)$.} 
		\label{d4cep}
\end{figure*}

Figure \ref{d4cep1} shows the conditional pdf $f(d_4 ~|~ d_4 > 36)$ and its expected value  41.07.
Figure \ref{d4cep2} plots  the conditional expectation 
$\hat{d}_{4}(t)$ as a function of $t$. 
	
\begin{remark}[Residual phase duration]
Predicting the residual phase duration, i.e., the time remaining for the phase to change from red to green (or vice versa),  might be more useful in eco-driving applications.  Vehicles can control their acceleration based on this information in order to minimize the fuel consumption. Our algorithm provides the residual phase duration. For example, the residual time of $p_4$ at time  $t$, $r_{4}(t)$ is simply
\begin{equation}
r_{4}(t) = \hat{d}_{4}(t) - t.
\end{equation}
Figure \ref{d4cep3} shows $r_{4}(t)$ as a function of $t$. 
\hfill $\Box$
\end{remark}	

\begin{remark}
One striking feature is  that the residual time $r_4$ (in Fig. \ref{d4cep3})  suddenly \textit{increases} at $t = 35$ by about 2.5s, which may appear counter-intuitive.  For example, consider a  driver waiting for the left turn signal, phase $p_1$, to turn green which coincides with the end of $p_4$. The residual time is decreasing initially as one would expect because intuitively $t$ increasing suggests less time remaining in $p_4$. However, the driver will find that residual time suddenly extended by a few more seconds at $t=35$ before decreasing again. (The phenomenon is similar to  the experience of the `remaining time' to download a file.) If the residual time is revealed to the driver via a countdown clock as in \cite{Audi_Spat}, then the clock must make a backward jump at $t = 35$. By assuming that the prediction of the duration does not change with time, i.e.	$\mathbb{E}[d_4~|~ d_4 > t] -t = \mathbb{E}[d_4] - t$, the eco-driving control strategies in \cite{rakha_spat_2011} and \cite{barth_spat_2011}   rule out the realistic situation depicted in  Fig. \ref{d4cep3}. This can create significant problems in designing a control strategies for eco-driving.  Of course the complexity of Fig. \ref{d4cep3} disappears in the case of fixed-time signals, whose pdfs are delta-functions.  \hfill $\Box$
\end{remark}

\subsection{Confidence based prediction}
\label{sec-confidence}
The SPaT message  includes the confidence level for the prediction. We can use the empirical pdf to give a prediction with a given confidence bound.

\begin{figure*}[t!]
\centering
\subfloat[]{\includegraphics[width=0.31\textwidth]{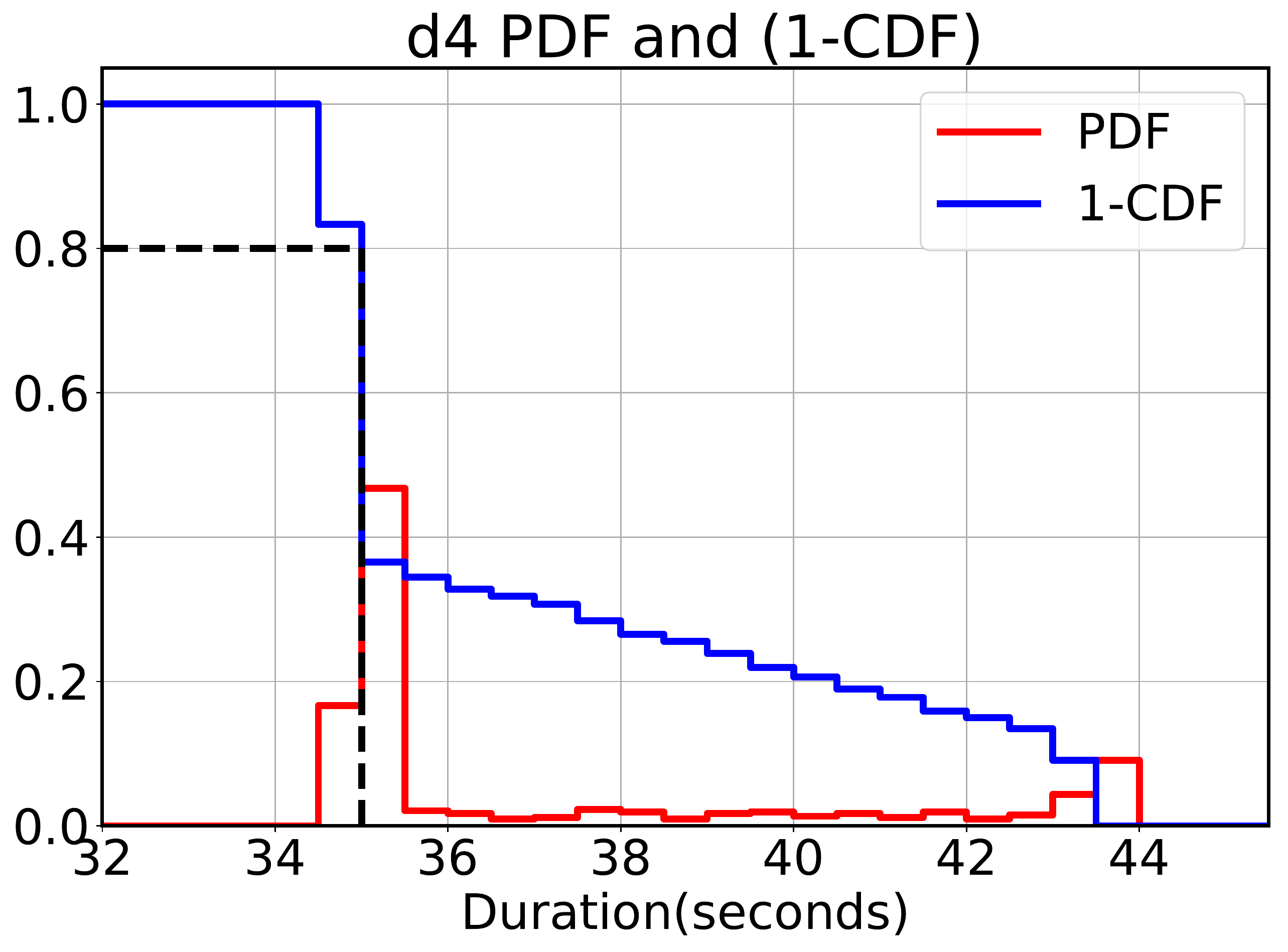}
\label{d4pdfcdf1}}
		\hfil
		\subfloat[]{\includegraphics[width=0.31\textwidth]{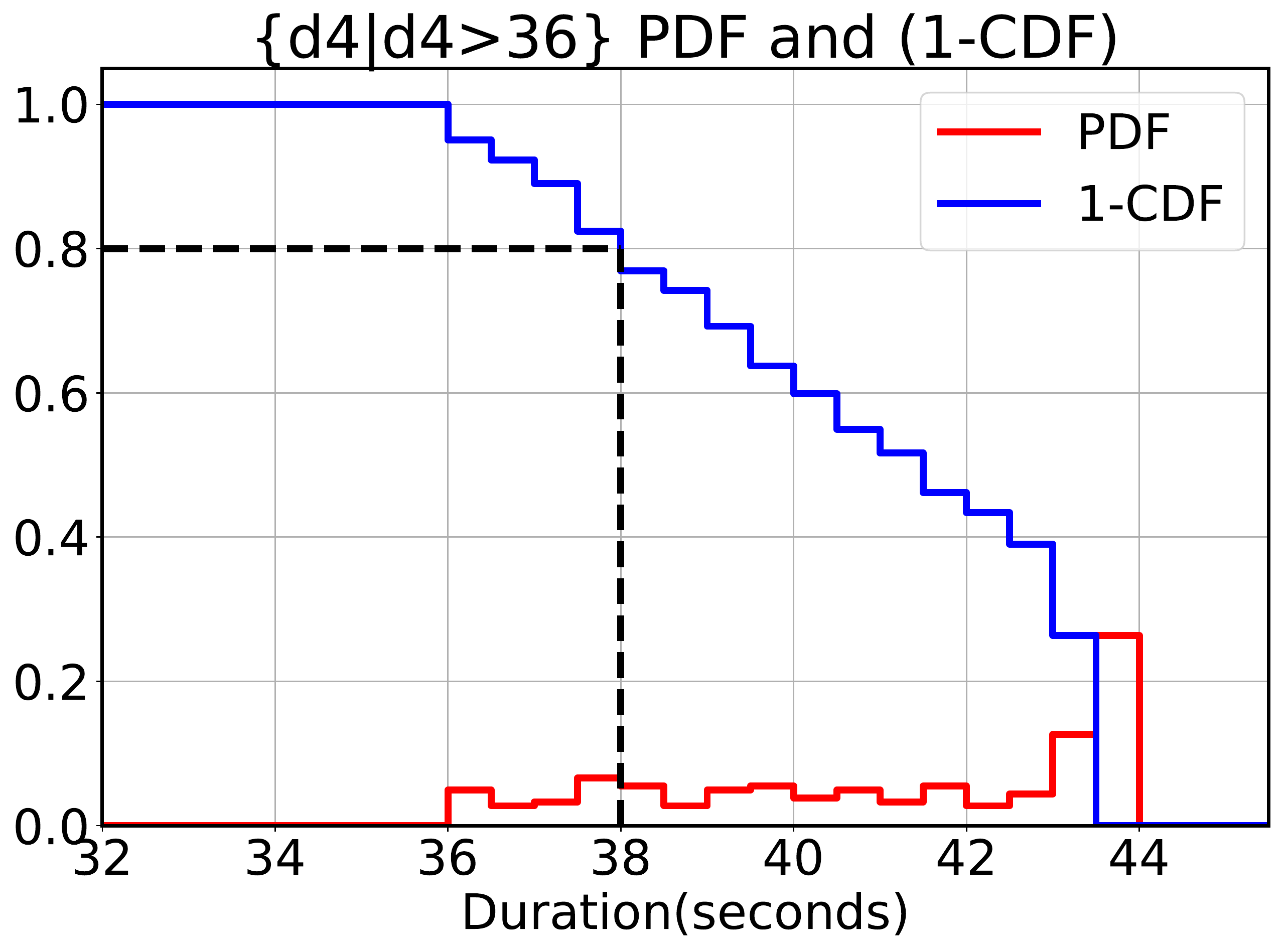}
			\label{d4pdfcdf2}}
		\hfil
		\subfloat[]{\includegraphics[width=0.31\textwidth]{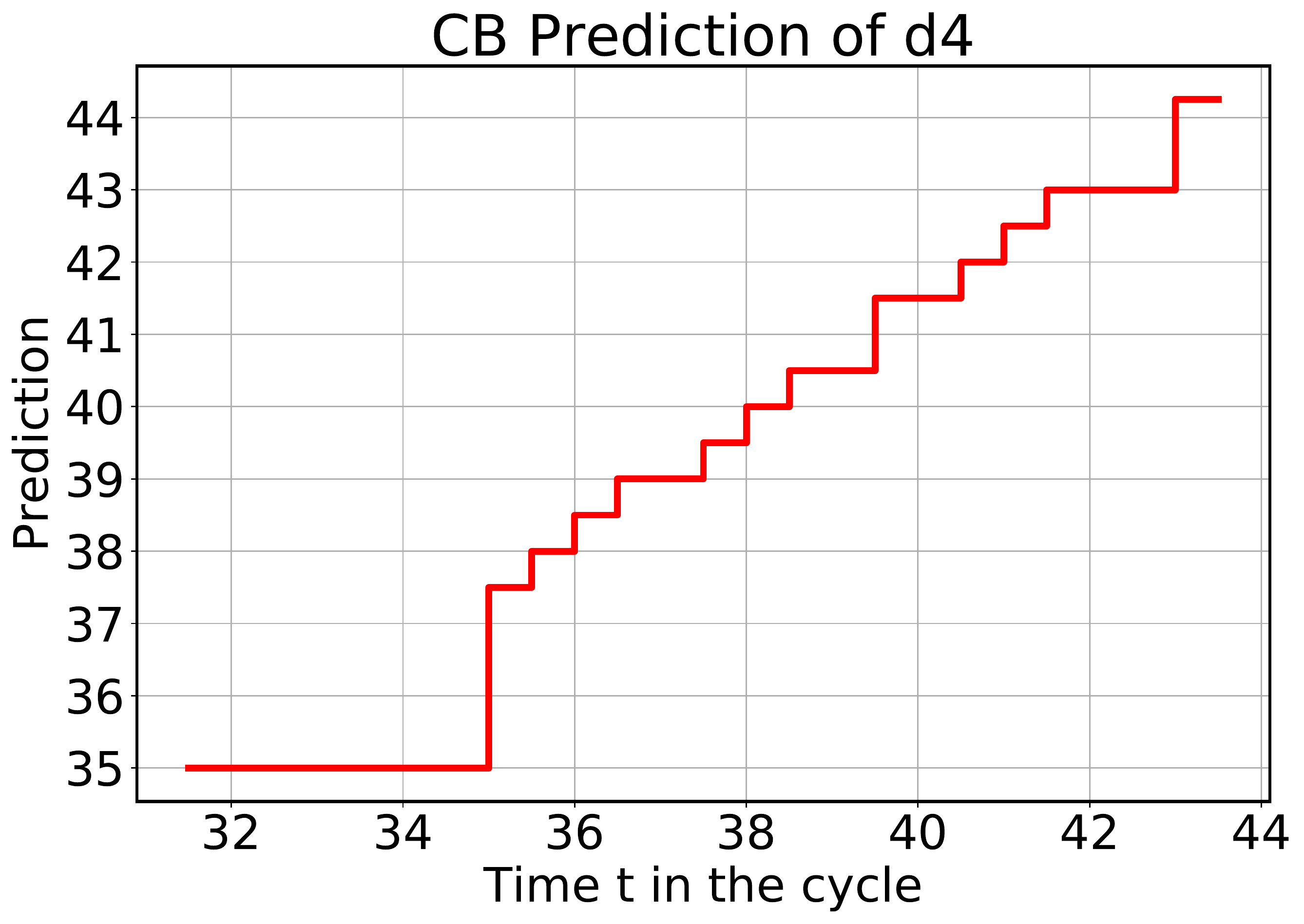}
			\label{d4cb1}}
		\caption{Confidence based prediction of $d_4$ for confidence bound $\alpha = 0.8$.}  
		\label{d4cb}
\end{figure*}

Let $\alpha$ be the required confidence bound. We can  define the confidence based prediction as the value $d$ for which one can  guarantee $\mathbb{P}(d_4 > d) = \alpha$.  Let $F(d) = \mathbb{P}(d_4 \leq d)$ be the cumulative distribution function (cdf) of $d_4$.  Then $1 - F(d) = \mathbb{P}(d_4 > d)$ and the required $d$ is the solution of the equation 
\begin{equation} 
\label{eq6}
1 - F(d) = \alpha.
\end{equation}
Fig. \ref{d4pdfcdf1} plots the pdf and 1-CDF of $d_4$. The latter is a decreasing function, and the solution of \eqref{eq6} is $d = 35$ for $\alpha=0.8$.  Now suppose we seek the confidence bound at time $t = 36$ into the cycle while $p_4$ is actuated.  Then the bound is given by 
\begin{equation} 
\label{eq7}
1 - F(d~|~d_4 > 36) = \alpha .
\end{equation}
Here $F(d~|~d_4 > 36)$ is the CDF of $d_4$ conditioned on the event $\{d_4 > 36\}$.  So $d = 38$ for $\alpha=0.8$ as shown in  Fig. \ref{d4pdfcdf2}. 

Thus with probability 0.8, at time $t =0$ in the cycle $d_4$ is at least 35, and at time $t = 36$ in the cycle $d_4$ is at least 38. Fig. \ref{d4cb1} plots  this confidence based prediction  as a function of $t$  for $\alpha=0.88$. 

\subsection{Prediction errors} 
\begin{figure*}[t!]
		\centering
		\subfloat[]{\includegraphics[width=0.31\textwidth]{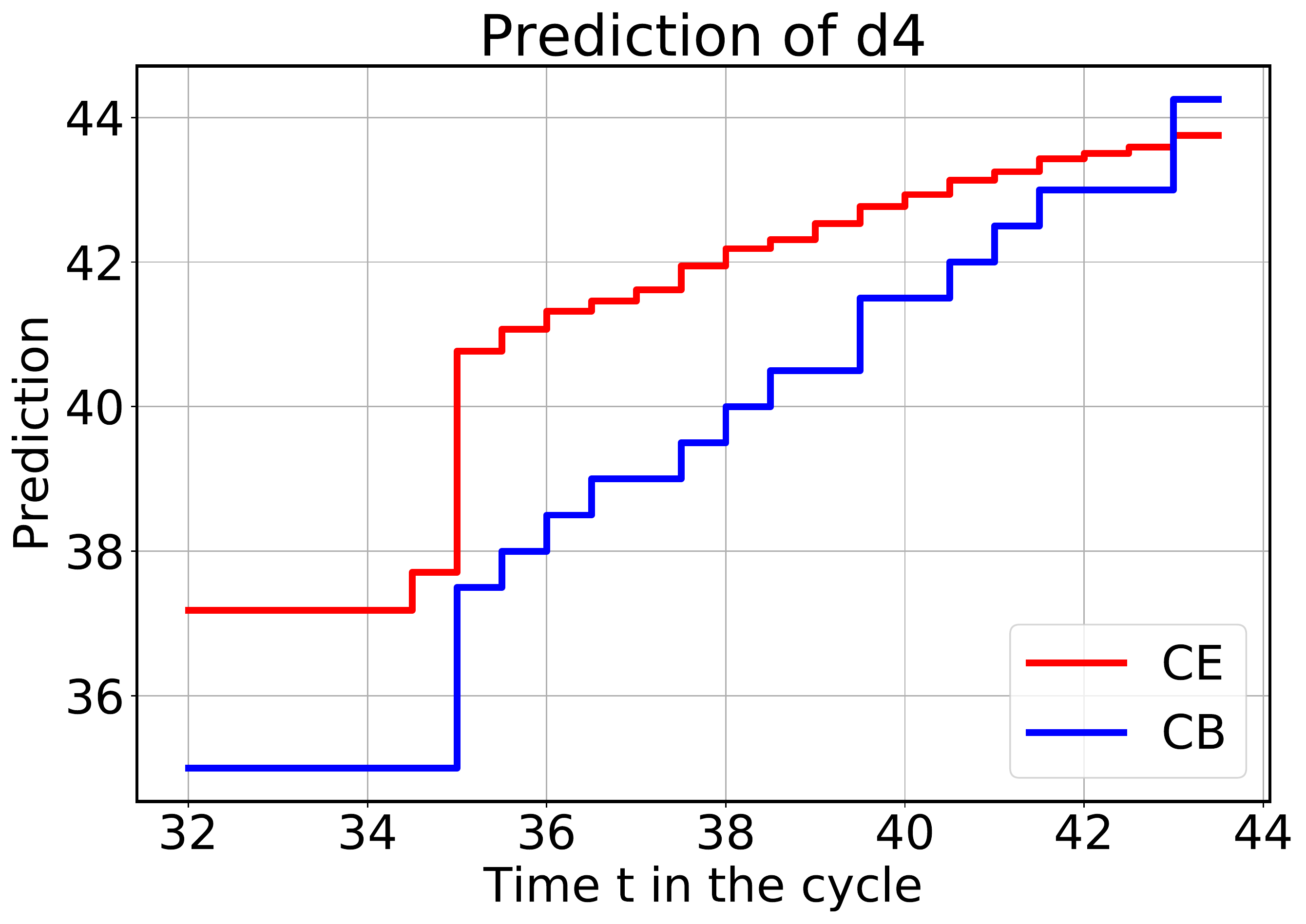}
			\label{d4cecb1}}
		\hfil
		\subfloat[]{\includegraphics[width=0.31\textwidth]{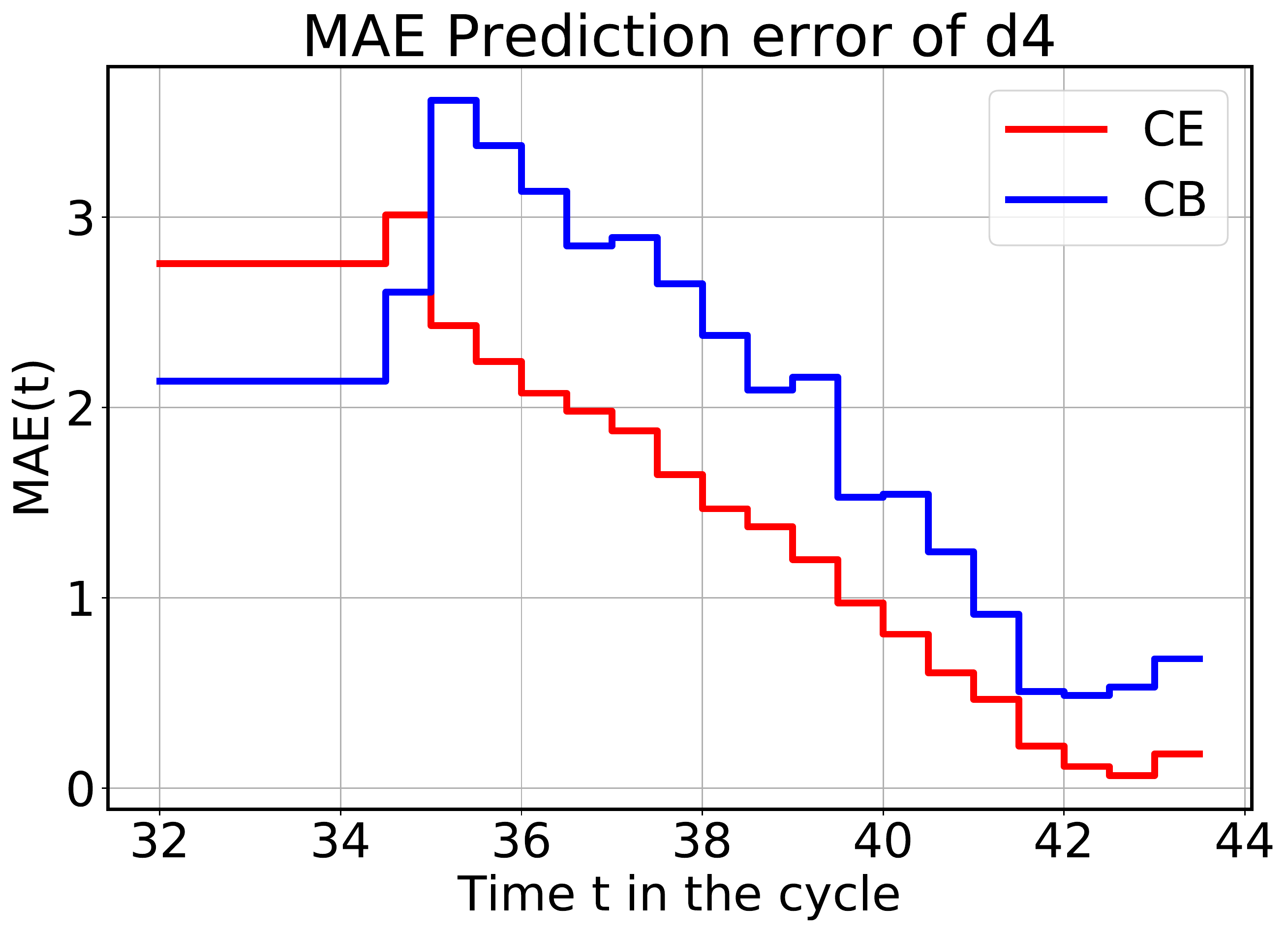}
			\label{d4mae}}
		\hfil
		\subfloat[]{\includegraphics[width=0.31\textwidth]{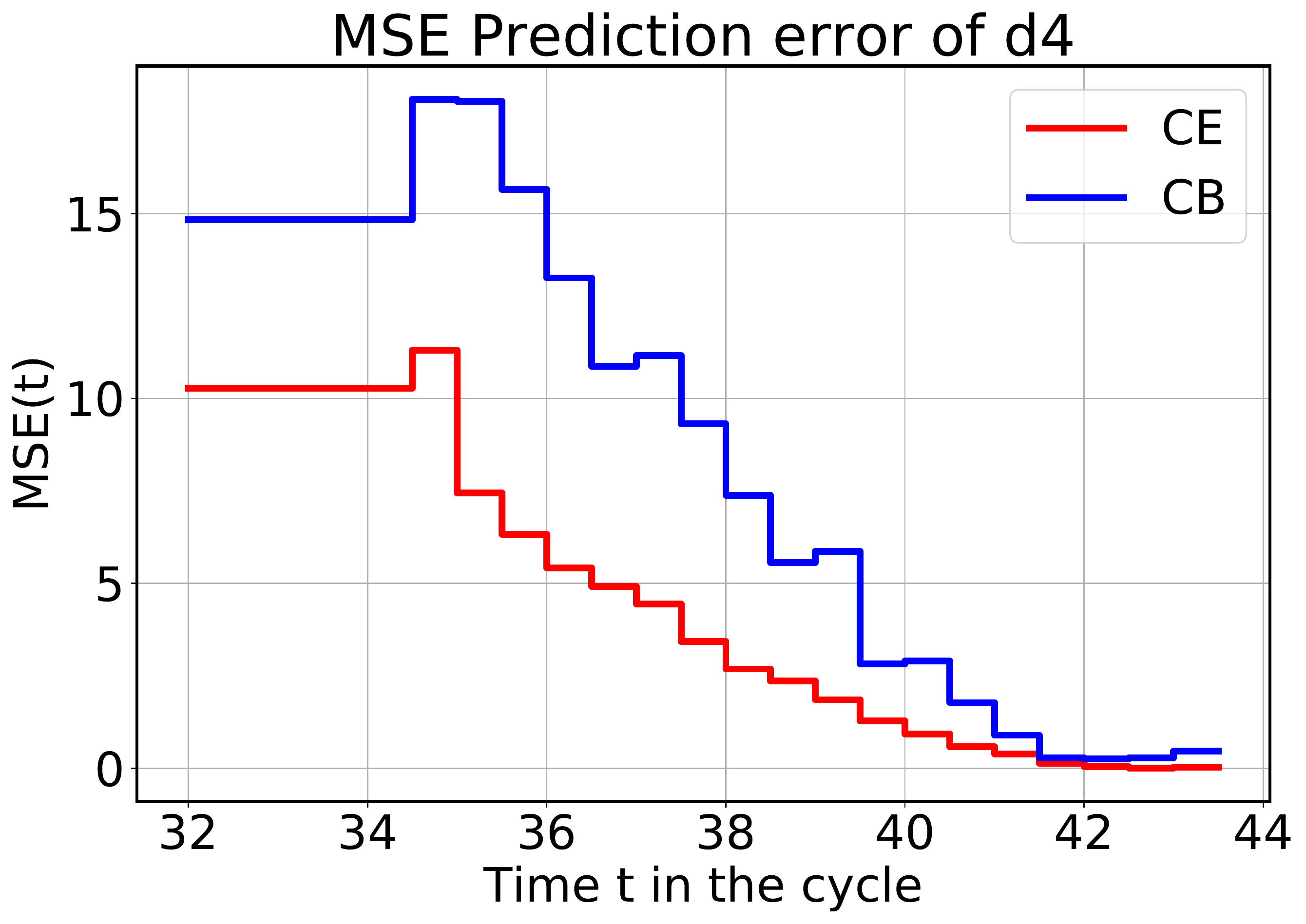}
			\label{d4mse}}
		\caption{Prediction error comparison.} 
		\label{k}
\end{figure*}
	
Consider the conditional expectation based prediction shown in Fig. \ref{d4cep2}   at time $t$ into the cycle, for which the prediction of the residual time is $\hat{d}_{4}(t)$ (For example, for $t = 42$, $\hat{d}_{4}(t) = 43.5$.)  In our data set there are many cycles or samples $\omega$ for which $d_4 (\omega) > t$ and for each  $\omega$   the exact error in the prediction of the residual time is $\hat{d}_{4}(t) - d_4(\omega)$.  So the mean absolute error $MAE(t)$ for the prediction at $t$ is
\begin{equation} \label{eq8}
MAE(t) = \frac{1}{n(t)} \sum_{\omega = 1}^{n(t)} |\hat{d}_{4}(t) - d_4 (\omega)|,
\end{equation}
in which $\omega = 1, ... , n(t)$ are the samples with $d_4 (\omega) > t$. 

\begin{remark}
Formula \eqref{eq8} may be  considered suspect because the calculation of $\hat{d}_{4}(t)$ includes the value of the sample $d_4(\omega)$. The `leave-one-out' error calculation replaces $\hat{d}_{4}(t)$ in  \eqref{eq8} by $\hat{d}_{4}(t, \bar {\omega})$ which is the prediction obtained after leaving out the sample $\omega$.  However, since the number of  samples is large this will not materially affect the prediction error. 
\hfill $\Box$
\end{remark}

In exactly the same way we can calculate the MAE for the confidence based  prediction shown in Figure \ref{d4cb1}.  The formula	for MAE is  the same as  \eqref{eq8}, the only difference is that $\hat{d}_{4}(t)$ is replaced by the confidence based estimate. Fig. \ref{d4cecb1} plots the conditional expectation based prediction and confidence based prediction for comparison. Figure \ref{d4mae} plots the $MAE(t)$ for both the  conditional expectation and confidence based predictions (with $\alpha=0.8$).  The most noteworthy aspect of the figure is that both prediction errors generally decrease as more real-time information is accumulated by the intersection.  
	
If we take a quadratic loss function, \eqref{eq8} must be replaced by the mean-squared error \eqref{eq9}:
\begin{equation} 
\label{eq9}
MSE(t ) = \frac{1}{n(t)} \sum_{\omega = 1}^{n(t)} [d_{4}(t) - d_4 (\omega)]^2.
\end{equation}
	in which $\omega = 1, ... , n(t)$ is the same as in \eqref{eq8}.  It is well known that the best prediction which minimizes the MSE is conditional expectation as confirmed in Fig. \ref{d4mse}.

Evidently, having real-time information significantly improves the prediction, both in the form of conditional expectation (Figure \ref{d4cep}) and with confidence bounds (Figure \ref{d4cb}).

\section{Prediction as Minimizing a Loss Function} 
\label{sec-loss}
Different predictors minimize different loss function. Let $x$ be the prediction and let $d$ be the actual realization of the the phase duration. The optimal prediction $d^{*}(t)$ at time $t$ is defined as
\begin{equation}
d^{*}(t) = \arg \min_{x} \mathbb{E}[\ell(x - d)~|~d>t ],
\end{equation}
where $\ell(\cdot)$ is the specific loss function we are considering and expectation is w.r.t. the random variable $d$. For example, while  considering the MSE \eqref{eq9}, the   loss function  used is $l(y) = y^{2}$.

The function $MAE$ in \eqref{eq8}  takes the loss  to be the absolute value of the error, i.e., $\ell(y) = |y|$. So a positive error and a negative error of the same magnitude are judged equally harmful.  But it seems more likely that an overestimate of the `time to red' is evaluated by a driver differently from the same error in the estimate of `time to green'. (In the former case, the driver may be forced to slam on the brakes.)  This consideration suggests using the asymmetric loss function 
\begin{equation}
\label{eq10}
\ell(y) = \left\lbrace \begin{array}{cc}
c_{1} |y| &\quad \text{if}~y<0 \\
c_{2} ~y &\quad \text{if}~y\geq0
\end{array} \right. .
\end{equation}

It is not  difficult to show that the optimal prediction $d^{*}(t)$ which minimizes \eqref{eq10}  is given by the formula
\begin{equation}
\label{eq11}
F(d^{*}(t)~|~d>t ) = \frac{c_{1}}{c_1 + c_2},
\end{equation}
where $F(\cdot ~|~d>t )$ is the conditional cumulative distribution. So this is just the confidence based estimate (like \eqref{eq6})  for an appropriate choice of the confidence bound.  

%

\begin{figure}[h!] 
		\centering
		\includegraphics[width=0.4\textwidth]{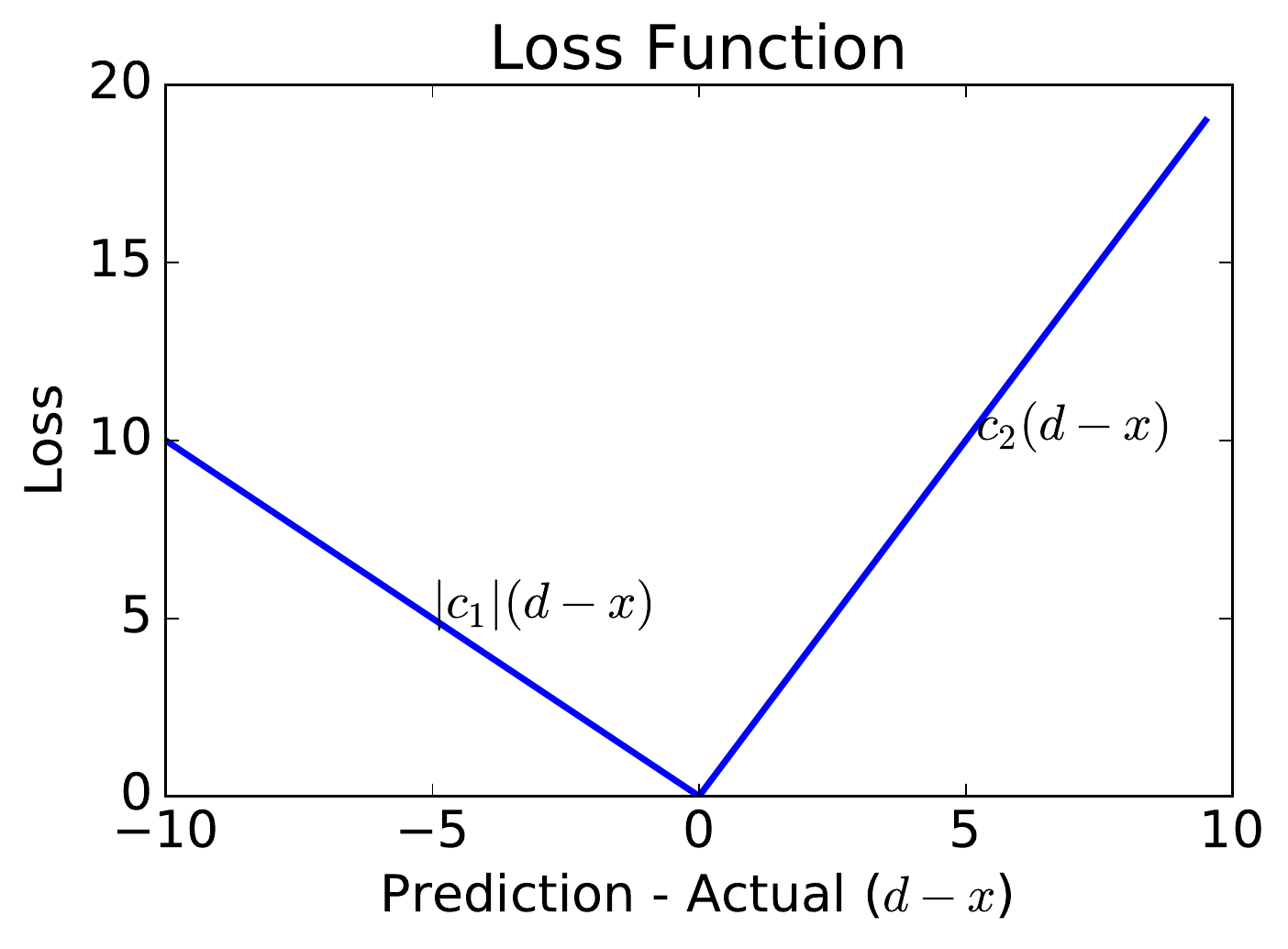}
		\caption{An asymmetric loss function in which the unit loss of an overestimate of the residual duration is different from the unit loss of an underestimate.} \label{fig10}
\end{figure}
This loss function is plotted in Figure \ref{fig10}.  If $|c_1| \neq c_2$, the two slopes are different.  So, if our driver feels 	$|c_1| < c_2$ when driving during phase $p_2$, he will feel the two slopes should be exchanged when driving during phase $p_4$. This suggests that the intersection may broadcast different predictions of the same phase duration for different approaches.

\section{Prediction of Other Phases} 
\label{sec:otherphases}

In Section \ref{sec-residual}  we considered prediction of $p_4$ in detail. We now consider a more complex prediction.  Suppose a driver  at time $t$ on Tildenwood Dr. (see Figure \ref{fig1}), while the minor phase $p_4$ is active, wants to know when the through phase $p_2$ will start.  We see from Figure \ref{fig3} that this means  the driver wants a prediction of $d_4 + d_1$ at time $t < d_4$.  We consider two approaches to an answer.
	
\textit{Approach 1:}  We treat $d = d_4 + d_1$ as a single random variable, and recognize that $d_4 > t $ implies $d > t$.	We obtain the empirical pdf of $d = d_4 + d_1$ and  we predict the residual duration as the conditional expectation
\begin{equation}\label{eq13}
	\mathbb{E}[d_4 + d_1 ~|~d_4 + d_1 > t].
\end{equation}

We can also calculate the prediction with confidence level $\alpha$ as the residual $d$ that solves the equation
\begin{equation}\label{eq14}
	1 - F(d~|~(d_4 + d_1 > t)) = \alpha. 
\end{equation}

\textit{Approach 2:} We treat $(d_4, d_1)$ as a 2-dimensional random variable, obtain the empirical conditional distribution 
	$\mathbb{P}(d_4 + d_1~|~ d_4 > t)$, and  then calculate the prediction as the conditional expectation
	$\mathbb{E}[d_4 + d_1~|~d_4  > t]$.  
	
	Approach 1 is approximate, Approach 2 is exact but requires calculating the two-dimensional probability distribution $f(d_4, d_1)$.

\begin{figure} [h!] 
		\centering
		\includegraphics[width=0.50\textwidth]{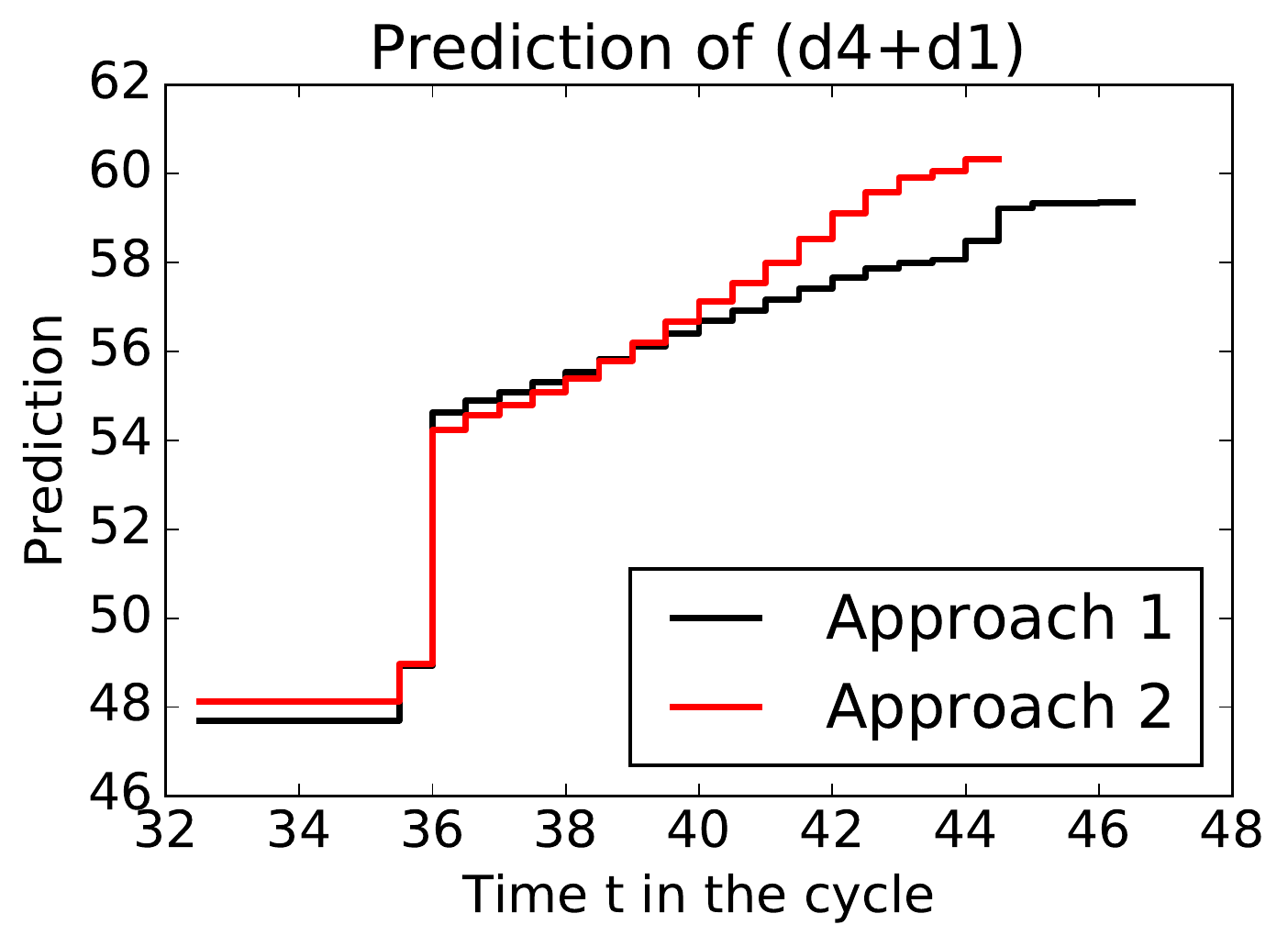}
		\caption{Conditional expectation based prediction of $(d_{4}+d_{1})$.}
		\label{d4d1pred}
	\end{figure}

\begin{figure} [h!] 
		\centering
		\includegraphics[width=0.50\textwidth]{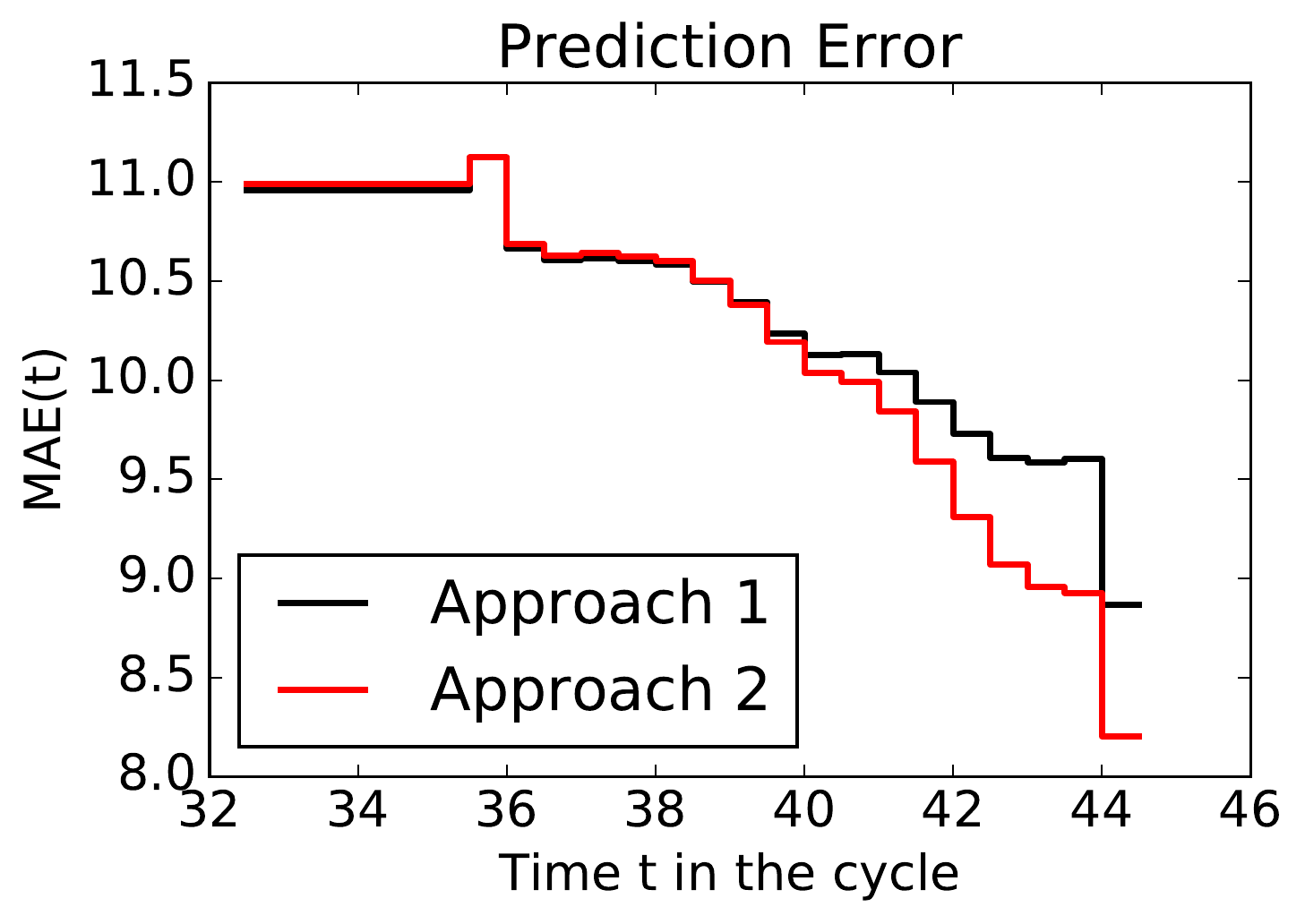}
		\caption{Prediction error of $(d_{4}+d_{1})$.}
		\label{d4d1prederr}
	\end{figure}

	The predictions following the two approaches  is plotted in Fig. \ref{d4d1pred}, and prediction error is plotted in Fig. \ref{d4d1prederr}. As expected, the prediction following Approach 2 has a smaller mean absolute error.

\section{Evaluation  at a Different Intersection}
\label{sec:site2}

 \begin{figure}[h!]
\centering
\includegraphics[width=0.45\textwidth]{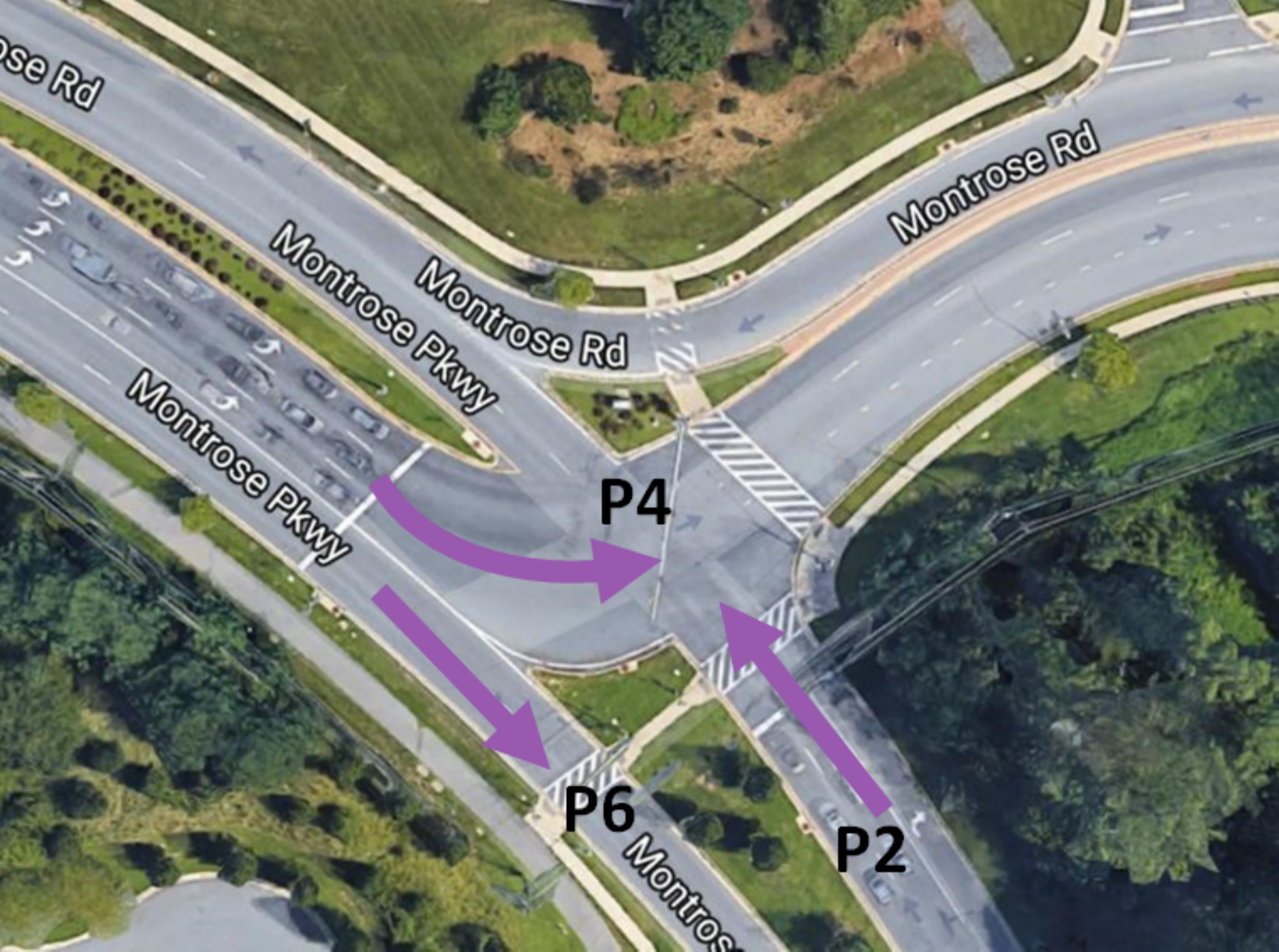} 
\caption{The intersection at Montrose Rd. and Motrose Pkwy.} 
\label{fig:site2}
\end{figure}

In this section we apply  our prediction algorithm to the intersection at Montrose Road and Montrose Pkwy shown in Figure \ref{fig:site2}.  The   figure also indicates the three phase movements ($p_{4}, p_{2}, p_{6}$) permitted at this intersection.  Each cycle  starts with the phase $p_{4}$ followed by phase $p_{2}$. The analysis here uses nine months of data from April 2016 to April 2017 (three months are omitted due to problems in getting data). 

\begin{figure*}[t!]
		\centering
		\subfloat[]{\includegraphics[width=0.31\textwidth]{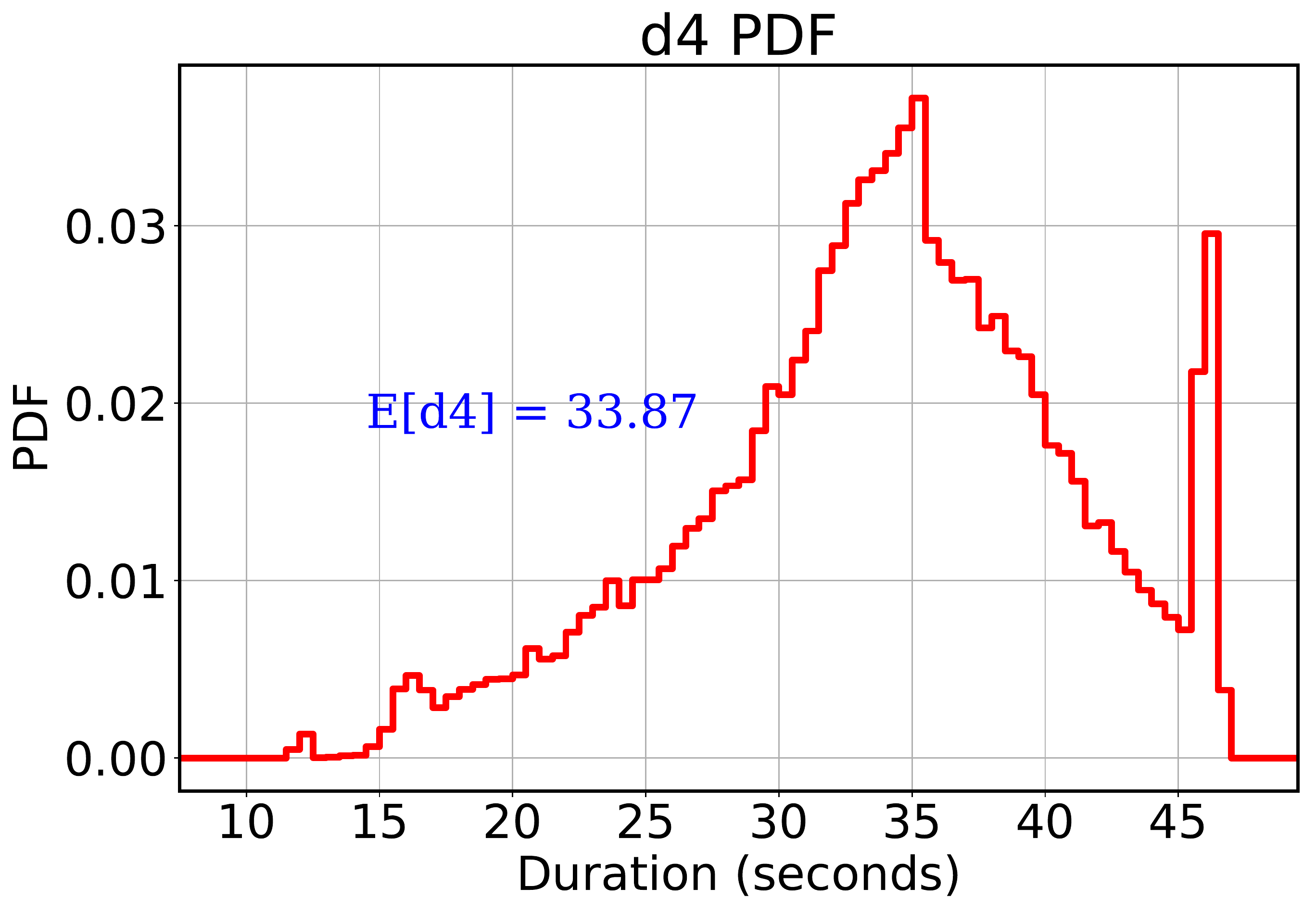}
			\label{d4pdf_2}}
		\hfil
		\subfloat[]{\includegraphics[width=0.31\textwidth]{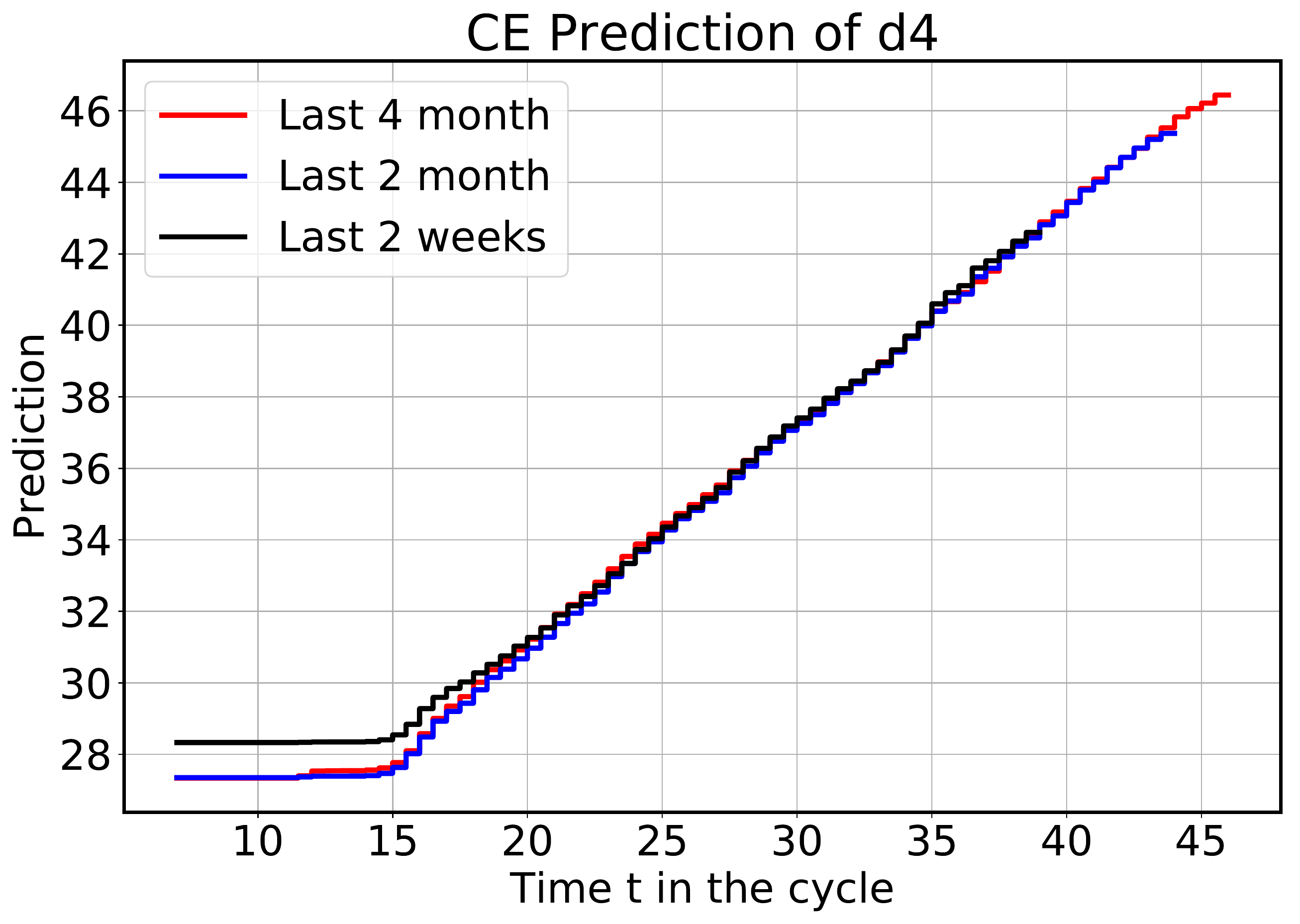}
			\label{d4ceppred_2}}
		\hfil
		\subfloat[]{\includegraphics[width=0.31\textwidth]{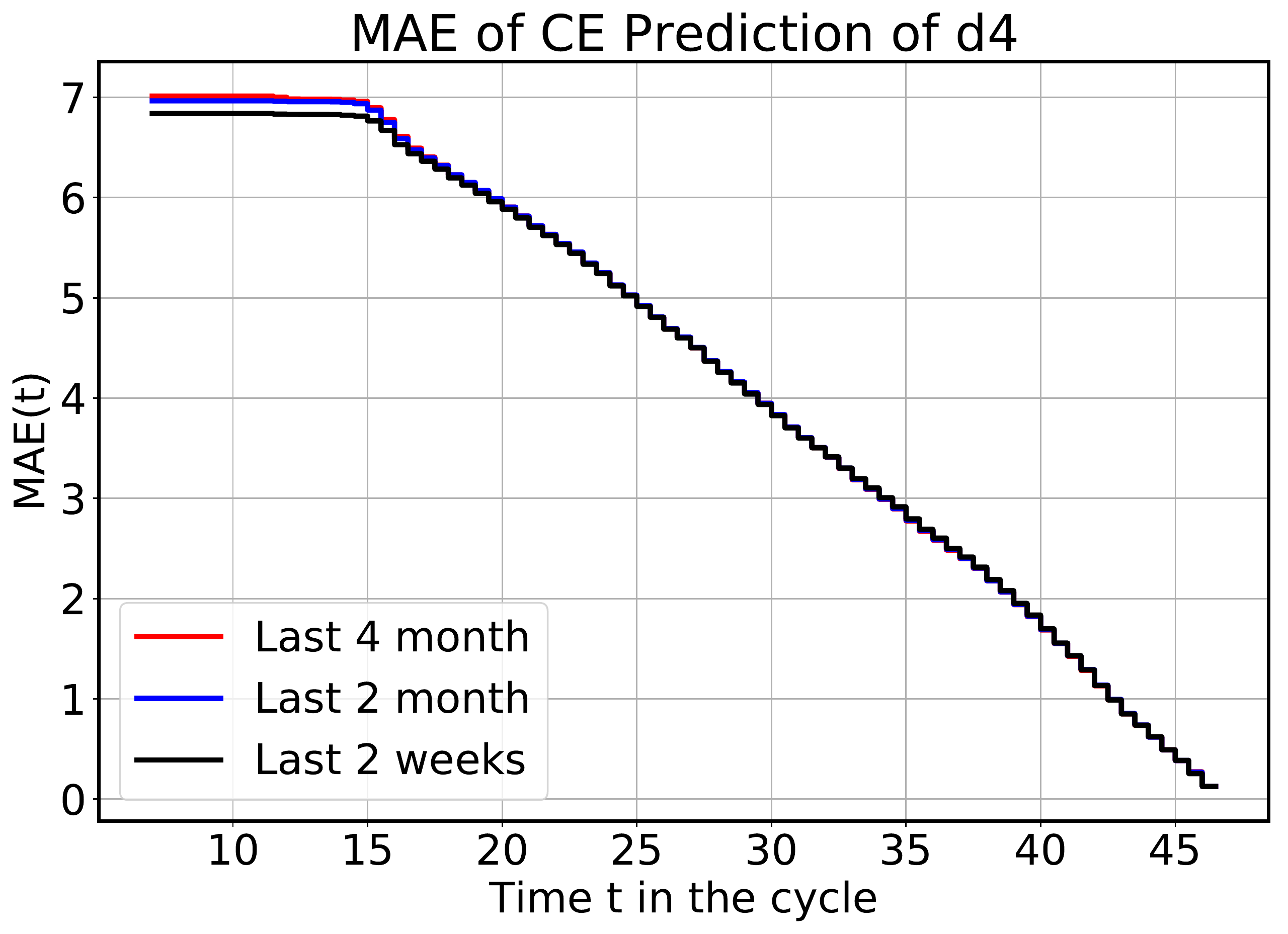}
			\label{d4cepmae_2}}
		\caption{Conditional Expectation based prediction for $d_{4}$.} 
		\label{k}
\end{figure*}

We focus on the left turn phase $p_{4}$.  The empirical pdf of $d_{4}$ shown in Fig.  \ref{d4pdf_2}  uses data for all   nine months. Note that the phase duration here is more variable compared to that of the earlier intersection.  

In order to get a sense of how much data is necessary to make the prediction (i.e. for computing the empirical pdf), we propose a \textit{sliding window} approach. For predicting the phase duration on day $n$, we use the data from the days $n-\Delta$ to $n-1$, where $\Delta$ is 14 days, 60 days, or 120 days. So, for each day, we compute  three different empirical pdfs and make three different prediction. The prediction is done in the  way  described in Section \ref{sec-conditional}. The prediction for one day using three different $\Delta$'s are given in Fig. \ref{d4ceppred_2}. The MAE averaged over the whole data is shown in Fig. \ref{d4cepmae_2}. The MAE is larger compared to the previous intersection because the phase duration is more variable.

\begin{remark}
The MAE plot (Fig. \ref{d4cepmae_2} ) shows that  the prediction error does not decrease just by using more data, because the  traffic, and hence the phase duration process, is not strictly stationary. The MAE is  similar for all  three window lengths (2 weeks, 2 months and 4 months).  If  the window length is smaller than 2 weeks, the prediction error increases.   \hfill $\Box$
\end{remark}	

We also evaluated the confidence based prediction (as described in Section \ref{sec-confidence}) for this intersection and obtained similar results. 
	

\section{Conclusions} \label{sec-conc}
	The paper describes  several algorithms for SPaT predictions, i.e. estimates of the remaining duration of  a signal phase.  The algorithms use historical and real-time phase data for a semi-actuated intersection in Montgomery County, MD.   The algorithms can be readily implemented at the intersection's signal controller.  We summarize three principal findings.
	
	First,  knowing how much time into the current phase has elapsed greatly improves the prediction of the residual time for that phase as well as for a subsequent phase.  Further, the important statistic underlying all the algorithms is the  probability distribution of the residual time conditioned on the available information.  This conditional probability is easy to compute and from the distribution one can readily construct the SPaT message.   For example, for the intersection of Figure \ref{fig1} the SPaT message for phase $p_4$ at time $t=0$ is: 
	
\hangindent=0.3in
\hspace*{0.28in}
	startTime = 0,
minEndTime = 35  (Fig \ref{d4pdf1}),
maxEndTime = 45.5 (Fig \ref{d4pdf1}),
likelyTime = 38.13 (Fig \ref{d4pdf1}),
confidence bound = 35 at $\alpha =0.8$ (Fig \ref{d4pdfcdf2}),
nextTime = 120 (cycle length).

	Second, for an actuated signal,  as  time increases, the real-time prediction of the residual time can increase as well as decrease.  This poses a challenge to the design of speed profiles that reduce fuel consumption.  Third,  drivers are likely to weight differently  errors in predicting `end of green' and `end of red', so drivers on two different approaches would prefer different estimates of the residual time of the same phase.  It may therefore be worth considering providing multiple estimates of the residual time.
	
Several issues  warrant further study.  Since the duration of a phase in an actuated intersection depends on vehicle detections during the same cycle, incorporating these in the estimate of the phase duration should reduce error.   In a future paper we will report the improvement in predictive power from vehicle detection measurements at the  intersection in Figure \ref{fig1}.  Second, Figure \ref{fig4}
	suggests that the sequence of phase durations is not stationary, and so one should not  estimate the empirical probability distributions by averaging over all available data.  Rather more recent data samples should be given more weight.  A more challenging approach is to try to estimate a model of the dependency among subsequent phase durations.  To the extent that this is possible, it would certainly improve the SPaT estimates.

\section*{Acknowledgment}
We are grateful to Zahra Amini, Sam Coogan, Christopher Flores,  Mike Hui, and Ardalan Vahidi for their comments.

\bibliographystyle{IEEEtran} 
\bibliography{SPaT-Refs}

\begin{IEEEbiography}
[{\includegraphics[width=1in,height=1.25in,clip,keepaspectratio]{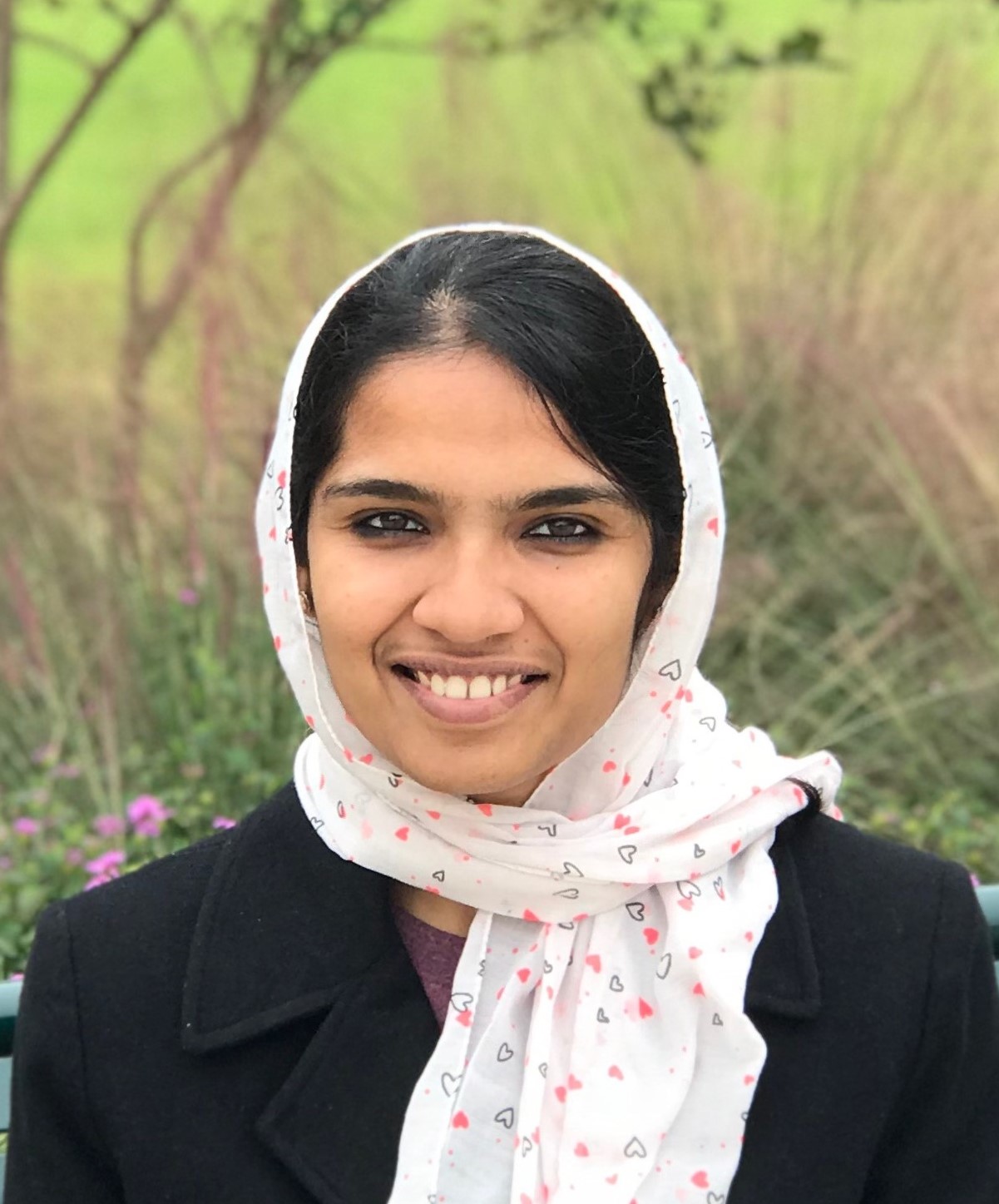}}]
{Shahana Ibrahim} is a graduate student in the  Department of Electrical and Computer  Engineering at Texas A\&M University.  She completed her Bachelors degree from National Institute of Technology, Calicut, India in 2012. Prior to starting her graduate education, she had been working as System Validation Engineer for wireless infrastructure products at Texas Instruments, Bangalore India from 2012 to 2017. Her research interests include in the broad areas of signal processing, machine learning and optimization. 
\end{IEEEbiography}

\begin{IEEEbiography}[{\includegraphics[width=1in,height
			=1.25in,clip,keepaspectratio]{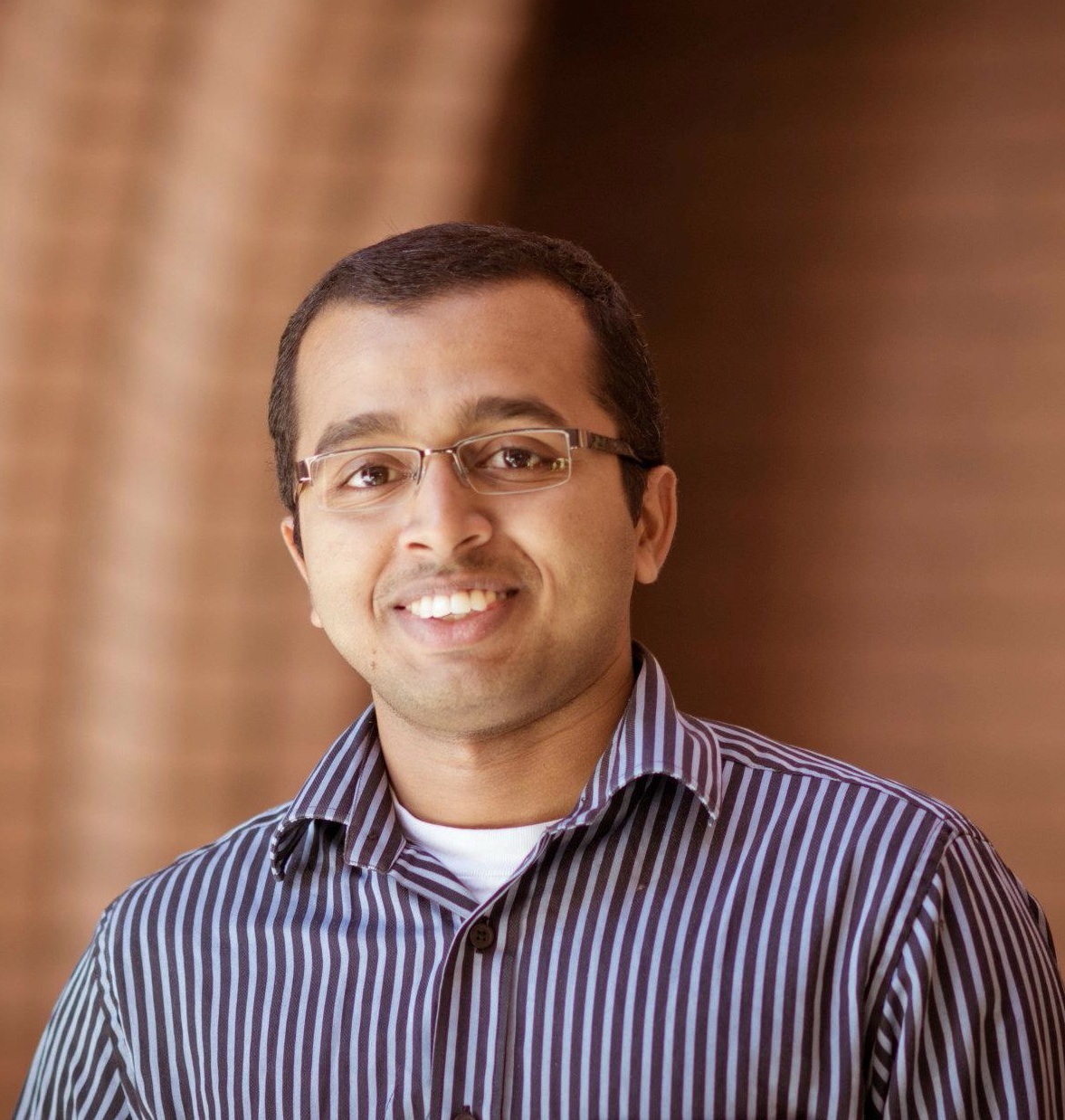}}]{Dileep  Kalathil}
		is an Assistant Professor in the Electrical and Computer Engineering Department at Texas A\&M University (TAMU) since August 2017. Before joining TAMU, he was  postdoctoral scholar in the Department of Electrical Engineering and Computer Sciences at the University of California, Berkeley. He received his PhD from University of Southern California (USC) in 2014 where he won the best PhD Dissertation Prize in the USC Department of Electrical Engineering. He received an M.Tech from IIT Madras where he won the award for the best academic performance in the EE department. His research interests include  Intelligent transportation systems, sustainable energy systems, data driven optimization, online learning, stochastic control, and game theory.
	\end{IEEEbiography}

\begin{IEEEbiography}
{Rene O. Sanchez} is a Sr. Systems Engineer at Sensys Networks. He received his PhD from the Department of Mechanical Engineering at the University of California, Berkeley in 2012. He received his B.S from the University of California, San Diego in 2006, where he graduated Summa Cum Laude from the Department of Mechanical and Aerospace Engineering.  He has in-depth research and industry experience in ITS projects. His areas of expertise include control theory, civil systems, optimization, data processing, tool development, system validation, system integration and algorithms. He is currently working on wireless sensors applications in transportation infrastructure, where he focuses on software and application development, advance data analytics and smart cities.
\end{IEEEbiography}

	\begin{IEEEbiography}[{\includegraphics[width=1in,height=1.25in,clip,keepaspectratio]{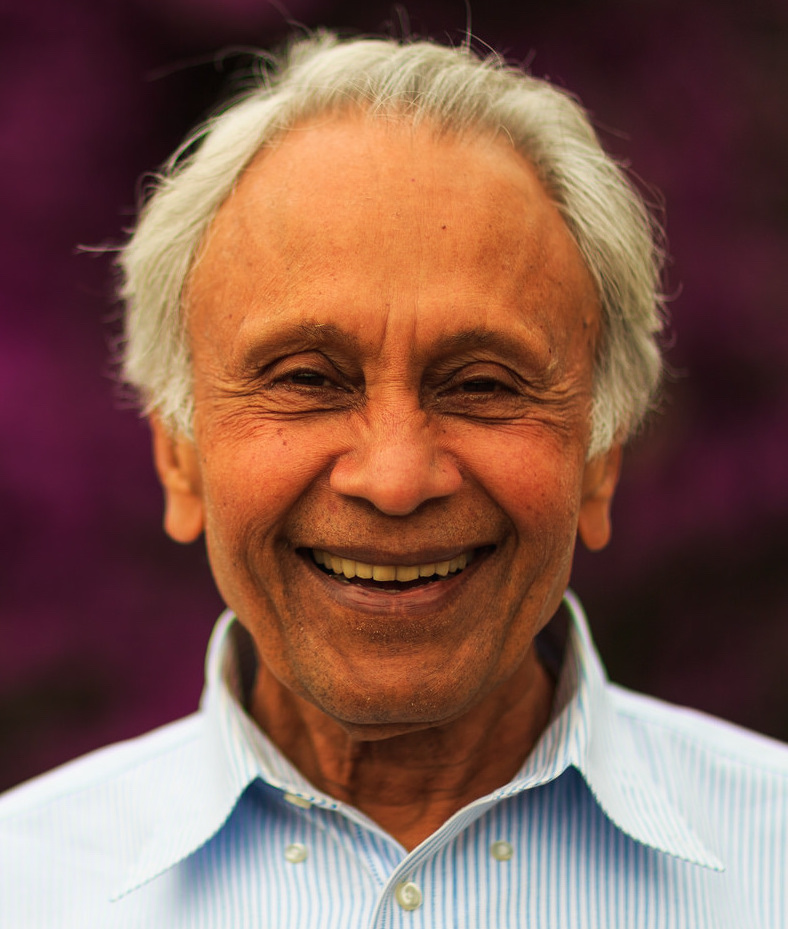}}]{Pravin Varaiya} is a Professor of the Graduate School in the Department of Electrical Engineering and Computer Sciences at the University of California, Berkeley.  He has been a Visiting Professor at the Institute for Advanced Study at the Hong Kong University of Science and Technology since 2010.  He has co-authored four books and 350+ articles.  His current research is devoted to electric energy systems and transportation networks.   
		
		Varaiya has held a Guggenheim Fellowship and a Miller Research Professorship.  He has received three honorary doctorates, the Richard E. Bellman Control Heritage Award, the Field Medal and Bode Lecture Prize of the IEEE Control Systems Society, and the Outstanding Researcher Award of the IEEE Intelligent Transportation Systems Society. He is a Fellow of IEEE, a Fellow of IFAC, a member of the National Academy of Engineering, and a Fellow of the American Academy of Arts and Sciences.
	\end{IEEEbiography}
	
\end{document}